\documentclass{emulateapj}
\usepackage[]{natbib, graphicx, float, appendix}
\usepackage[FIGTOPCAP]{subfigure}
\usepackage[section]{placeins}
\usepackage{wrapfig}
\usepackage[fleqn]{amsmath}
\def\vvel{\hbox{${\bf v}$}}

\newcommand{\RNum}[1]{\uppercase\expandafter{\romannumeral #1\relax}}

\begin{document}

\begin{abstract}

The evolution of the magnetic field from the large-scale dynamo is considered a central feature of the accretion disk around a black hole.  The resulting low-frequency oscillations introduced from the growth and decay of the field strength, along with the change in field orientation, play an integral role in the accretion disk behavior.  Despite the importance of this process and how commonly it is invoked to explain variable features, it still remains poorly understood.  We present a study of the dynamo using a suite of four global, high-resolution, MHD accretion disk simulations.  We systematically vary the scale height ratio and find the large-scale dynamo fails to organize above a scale height ratio of $h/r \gtrsim0.2$.  Using spacetime diagrams of the azimuthal magnetic field, we show the large-scale dynamo is well-ordered in the thinner accretion disk models, but fails to develop the characteristic ``butterfly" pattern when the scale height ratio is increased, a feature which is also reflected in the power spectra. Additionally, we calculate the dynamo $\alpha$-parameter and generate synthetic light curves.  Using an emission proxy, we find the disks have markedly different characters as stochastic photometric fluctuations have a larger amplitude when the dynamo is unordered.

\end{abstract}


\title{The influence of accretion disk thickness on the large-scale magnetic dynamo.}

\author{J.~Drew~Hogg\altaffilmark{1,2,3} and Christopher~S.~Reynolds\altaffilmark{1,2,4}}

\altaffiltext{1}{Department of Astronomy, University of Maryland, College Park, MD 20742, USA}
\altaffiltext{2}{Joint Space Science Institute (JSI), University of Maryland, College Park, MD 20742, USA}
\altaffiltext{3}{NASA Earth and Space Science Fellow}
\altaffiltext{2}{Joint Space Science Institute (JSI), University of Maryland, College Park, MD 20742, USA}
\altaffiltext{4}{Institute of Astronomy, Madingley Road, Cambridge CB3 0HA, UK}

\maketitle

\section{Introduction}
\label{sec-intro}

The accretion of gas onto compact objects remains a poorly understood astrophysical phenomenon.  For the standard thin accretion disk around a black hole, the magnetorotational instability \citep[MRI;][]{Velikhov59, 1960PNAS...46..253C, 1991ApJ...376..214B} is believed to be the chief mediator of angular momentum transport.  The MRI will quickly destabilize any weakly-magnetized rotating plasma with a Keplerian-like radial shear and drive turbulence.  Through the turbulence, correlated fluctuations in the fluid velocities (the Reynolds stress $R_{r \phi} = \rho v_{R} \delta v_{\phi}$) and correlated fluctuations in the magnetic field (the Maxwell stress $M_{r \phi} = -B_{r} B_{\phi}/4\pi$) \citep{1994MNRAS.271..197B} arise which produce a net internal stress that provides the kinematic viscosity considered by \citet{1973A&A....24..337S}.  The dimensionless ratio between the internal disk stress and gas pressure is \begin{equation}
\label{eqn-alpha}
\alpha_{SS}=\frac{\langle M_{r\phi} + R_{r\phi}\rangle}{\langle P \rangle}.
\end{equation} To sustain the magnetic field against dissipation, a mechanism must be present to rapidly regenerate field.

The vigorous MRI-driven turbulence and shear expected in an accretion disk provide conditions that make it prime territory for the development of a magnetic dynamo \citep{1992MNRAS.259..604T}.  Indeed, the self-organization of the magnetic field on large scales has been universally observed in simulations of moderately magnetized, thin accretion disks.  Despite the ubiquity with which dynamo behavior develops, the phenomenon remains an enigma and much of the fundamental theory behind its global growth remains undeveloped.

The large-scale dynamo often presents itself in simulations as a quasiperiodic reversal of the azimuthal field polarity.  Bundles of field rise as the overmagnetized regions feel a buoyant force, and spacetime diagrams show a characteristic ``butterfly pattern," akin to that observed in the migration of sunspots on the Sun.  This behavior is often interpreted as the evolution of the mean field within the framework of the $\alpha\Omega$ dynamo model, which has two ingredients.  First is the $\alpha$ effect which is sourced in the induction of azimuthal field from the movement of radial and the vertical magnetic field in a helical fluid flow \citep{1980mfmd.book.....K}.  This acts to generate large-scale field from small-scale turbulent motions.  Second is the $\Omega$ effect which arises through differential rotation and grows azimuthal magnetic field back from radial and vertical fields.  This seeds the MRI from the large-scale field and further sustains turbulence allowing the cycle to continue.  For the sake of clarity, we denote the $\alpha$ effect parameter as $\alpha_d$ henceforth to prevent confusion with the effective \citeauthor{1973A&A....24..337S} $\alpha$-parameter.

Shearing box simulations have been instrumental in allowing for the detailed exploration of the relation between the accretion flow and large-scale magnetic field, thus enabling the assembly of many of the pieces in the dynamo puzzle.  Early simulations demonstrated the sensitivity of the magnetic stresses to the net field spanning the domain \citep{1995ApJ...440..742H, 2004ApJ...605..321S, 2007ApJ...668L..51P} and that large-scale magnetic cycles with periods of roughly ten times the orbital period readily develop \citep{1995ApJ...446..741B, 2008A&A...488..451L, 2010ApJ...713...52D, 2011ApJ...728..130G, 2011ApJ...740...18O}.  Considering only a local patch of an accretion disk allows for the rigorous investigation of the structure of the turbulence, including its spectral properties \citep{2015ApJ...802..139M, 2017arXiv170707044G} and saturation \citep{2008A&A...487....1B, 2009MNRAS.394..715L, 2009A&A...498..241O, 2009ApJ...698L..72P, 2010ApJ...716.1012P}, which is important because the large-scale field grows from turbulent fluctuations.

Ultimately, though, the disk dynamo is a global feature and studying it requires models with large domains that properly account for the vertical and radial gradients in the accretion flow, as well as the coupling of radii with different evolutionary times.  Like their local counterparts, global models find dynamo periods of $10-20\times$ the local orbital period \citep[e.g.][]{2011ApJ...736..107O, 2011MNRAS.416..361B, 2012ApJ...744..144F, 2013ApJ...772..102H, 2013ApJ...763...99P, 2016ApJ...826...40H}.  These simulations are thin disks with typical thermal scaleheight ratios of $h/r=0.07-0.1$.  They find that the dynamo coherently spans roughly $\Delta r = 10\:r_g$ in radius, modulates the accretion disk stresses, and collects into sheets of azimuthal magnetic field of the same orientation in the coronal region.

In this paper, we aim to understand how the timing properties of the dynamo and large-scale magnetic field evolution depend on accretion disk geometry, i.e. the disk scaleheight ratio, and if there is a threshold of either large or small thickness beyond which the large-scale dynamo cannot be excited.  Exploring this is crucial because dynamos have been directly invoked to explain time variability signatures from accreting black holes \citep{2004MNRAS.348..111K, 2006MNRAS.368..379M}, as well as in several peripheral contexts.  Examples include dynamo cycles as a source of the low-frequency quasi-periodic oscillation \citep[QPO][]{2011ApJ...736..107O}, the driver of ``propagating fluctuations" in mass accretion rate \citep{2016ApJ...826...40H}, and as the trigger behind the secular evolution in the spectral state transitions \citep{2015ApJ...809..118B}.

Furthermore, while numerical simulations have established the robustness of the dynamo in a standard thin accretion disk, several cases have been found where the well-ordered oscillations are altered or vanish in atypical disk conditions.  For instance, using global simulations of a super-Eddington accretion flow, \citet{2014ApJ...796..106J} found that the dynamo period is regular, but slower.  The dynamo can be quenched in a magnetically dominated disk \citep{2013ApJ...767...30B, 2016MNRAS.457..857S} or if hydrodynamic convection acts to mix the field \citep{2017MNRAS.467.2625C}.  Additionally, the transition in the flow geometry of a truncated disk and its associated flow dynamics has been shown to impede the dynamo and lead to a sporadic, intermittent oscillation of the large-scale magnetic field in the inner hot disk \citep{2018ApJ...854....6H}.  Of course, a lingering concern is always that simulations are affected by nonphysical sensitivities like resolution and domain aspect ratios, which have also been shown to halt the dynamo \citep{2017arXiv170408636W}.

To delve into the scaleheight dependence of the magnetic dynamo, we constructed a suite of global, MHD disk models with varying scaleheight ratios.  In Section \ref{sec-model} we describe the numerical simulation of these models including the code details, initial conditions, and resolution properties.  In Section \ref{sec-results} we present an analysis of the large-scale dynamo properties, measure $\alpha_{\rm d}$ values for each simulation, briefly look at the evolution of the large-scale helicity in one of our simulations, and discuss potential observational characteristics of each simulation.  We discuss our results and their broader implications in Section \ref{sec-discussion} and provide closing remarks in Section \ref{sec-conclusion}.

\section{Numerical Model}
\label{sec-model}

In this paper we consider four well-resolved MHD simulations of model accretion disks in a pseudo-Newtonian potential, Equation \ref{eqn-pnpot}, with scaleheight ratios: $h/r=\{0.05, 0.1, 0.2,0.4\}$.  The goal of this paper is to use these simulations to understand how the accretion disk geometry affects the properties of the large scale magnetic field.  Hence, we strive for consistency in our models and to evolve the models for long enough to fully sample the global dynamo for several cycles.  To this end, these simulations were initialized with the same initial conditions and each were allowed for evolve for $t\approx4\times10^{4}\:GM/c^3$, or roughly 650 ISCO orbits. 

\begin{deluxetable*}{cccccccccccc} 
\tabletypesize{\normalsize} 
\tablecolumns{12} 
\tablewidth{0pt} 
\tablecaption{ Simulation Parameters \label{table-sim_params}} 
\tablehead{ 
\colhead{Simulation} & \colhead{Total Duration} & \colhead{ISCO Orbits} & \colhead{$N_r$} & \colhead{$N_\theta$} & \colhead{$N_\phi$} & \colhead{$H/\Delta \theta$} & \colhead{$\langle Q_{\theta} \rangle$} & \colhead{$\langle Q_{\phi} \rangle$} & \colhead{$\langle \Theta_B \rangle$} & \colhead{$\langle \alpha_{SS} \rangle$} & \colhead{$\langle \beta \rangle$} }
\startdata 
\texttt{hr\_005} & $3.58 \times 10^4 \; GM/c^3$ & $581.5$ & $1232$ & $248$ & $256$ & $28$ & $15.9$ & $37.7$ & $12.3$ & $0.065$ & $9.1$ \\
\texttt{hr\_01} & $3.88 \times 10^4 \; GM/c^3$ & $630.5$ & $616$ & $248$ & $128$ & $28$ & $11.8$ & $30.1$ & $12.4$ & $0.057$ & $10.1$  \\
\texttt{hr\_02} & $4.19 \times 10^4\; GM/c^3$ & $680$ & $308$ & $248$ & $64$ & $28$ & $11.9$ & $28.7$ & $12.4$ & $0.052$ & $10.4$ \\
\texttt{hr\_04} & $4.15 \times 10^4\; GM/c^3$ & $675.5$ & $154$ & $248$ & $32$ & $28$ & $5.7$ & $16.6$ & $11.8$ & $0.026$ & $11.8$

\enddata 
\end{deluxetable*}

\subsection{Simulation Code}

This work uses the second-order accurate PLUTO \emph{v4.2} code \citep{2007ApJS..170..228M} to solve the equations of ideal MHD in conservative form, \begin{eqnarray}
\frac{\partial \rho}{\partial t}+\nabla\cdot(\rho\vvel)&=&0,\\
\frac{\partial}{\partial t}(\rho\vvel)+\nabla\cdot(\rho\vvel\vvel-\bf{BB}+P{\cal I})&=&-\rho\nabla\Phi, \qquad \\
\frac{\partial}{\partial t}(E+\rho\Phi)+\nabla\cdot\left[(E+P+\rho\Phi)\vvel-\bf{B}(\vvel\cdot\bf{B})\right]&=&-\Lambda,\\
\frac{\partial{\bf{B}}}{\partial{t}}={\bf\nabla\times}({\bf{v}\times{B}}),
\nonumber
\end{eqnarray}
where $\rho$ is the gas density, $\vvel$ is the fluid velocity, $P$ is the gas pressure, {\boldmath$B$} is the magnetic field, ${\cal I}$ is the unit rank-two tensor, $E$ is the total energy density of the fluid,
\begin{equation}
E=u+{1\over 2}\rho |\vvel|^2 +\frac{\bf{B}^2}{2},
\end{equation}  
and $\Lambda$ accounts for radiative losses through cooling.  All fluid variables (e.g. $\rho$, $P$, $T$) are evolved and reported in a scale-free, normalized form.  The {\tt hlld} Riemann solver was used to solve the MHD equations in the dimensionally unsplit mode.  Linear reconstruction is used in space and the second-order Runge Kutta algorithm is used to integrate forward in time.  As a Godunov code, PLUTO conserves the total amount of energy in the simulation except for losses across the boundary and energy removed through the cooling function.  To enforce the $\nabla \cdot {\bf B} = 0$ condition, the method of constrained transport is used.

\subsection{Simulation Setups}

To remain as consistent as possible between models, the simulation grids are designed under the following strategy.  Each model is set in spherical coordinates with $\theta \in [\pi/2-5 h/r, \pi/2+5 h/r]$.  As we vary the scaleheight ratio, we keep the number of grid cells in this direction ($N_\theta=248$) the same.  Uniform grid spacing is used within $\pm 3\:h/r$ around the disk midplane with 28 zones per scaleheight.  A stretched grid is used beyond $3\:h/r$ with 40 zones in each of the coronal regions where there is less small scale structure. The only exception being the thickest $h/r=0.4$ disk, which we only model with $\theta \in [\pi/2-2.5\:h/r, \pi/2+2.5\:h/r]$ and halve the grid cells, accordingly.  The radial range stays the same between models ($r \in [5\:r_g, 145\:r_g]$) and is divided into an inner well resolved region used for the analysis ($r \in [5\:r_g, 45\:r_g]$) and an outer less resolved region ($r \in (45\:r_g, 145\:r_g]$) that acts as a gas reservoir to mitigate artificial effects from draining of the disk material that could introduce secular changes in the accretion flows.  Logarithmic spacing is used to keep $\Delta r / r$ constant along the grid.  The azimuthal ($\phi \in [0, \pi/3]$) domain is also fixed between models and has uniform spacing in $\phi$.  When varying the scaleheight ratio by factors of 2 between models, we change the number of radial and azimuthal grid zones to preserve the cell aspect ratio ($\Delta r: \Delta r \sin{\theta} \Delta \theta: r\Delta \phi \approx 1:1:2$).  This helps remove any resolution dependencies that could influence the results and adjusts the integration timestep, which is set by the Courant condition.  The validity of restricting the $\phi$-domain to $\phi \in [0, \pi/3]$ between our models rather than adjusting it in a similar manner to that used in the $\theta$-domain is addressed in Appendix \ref{sec-appendix}.  The full details of the simulation grids are given in Table \ref{table-sim_params}.  Outflow is allowed through the $r$ and $\theta$ boundaries while the $\phi$ boundaries are periodic.  Density and pressure floors are imposed to prevent artificially low or negative values.

A psuedo-Newtonian potential of the form,
\begin{equation}
\label{eqn-pnpot}
\Phi=-\frac{GM}{R-2 r_{\rm g}}, \qquad r_{\rm g}\equiv\frac{GM}{c^2}
\end{equation}
is used to approximate the dynamics of a general relativistic flow around a non-rotating black hole.  This captures features like the shear profile and the presence of an innermost stable circular orbit (ISCO) at $r=6\:r_{\rm g}$ without the computational expense of fully including general relativity.  A $\gamma=5/3$ adiabatic equation of state is used for the gas; the internal energy density of the gas is hence given by $u=P/(\gamma-1)$.  

\begin{figure}
\centering
\includegraphics{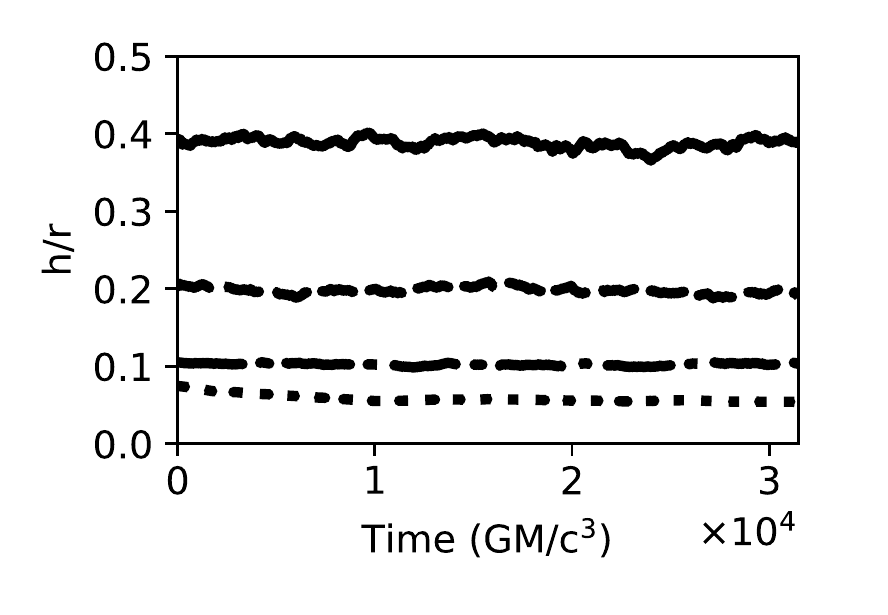}
\caption{Time variability of disk scaleheight ratios for \texttt{hr\_005} (dotted line), \texttt{hr\_01} (dot-dash line), \texttt{hr\_02} (dashed line), and \texttt{hr\_04} (solid line).
\label{fig-hr_time}}
\end{figure}

As the turbulence decays, it deposits energy in the gas in the form of heat so the disk would tend to become thicker.  To enforce the target disk aspect ratio for each of the simulations, a \citet{2009ApJ...692..411N} style cooling function is used to emulate radiative losses.  The local target gas pressure is set to,
\begin{equation}
P_{targ}= \frac{\rho v_{\rm K}^2 (h/r)^2}{\gamma},
\end{equation} where $v_{\rm K}$ is the local Keplerian velocity given by 
\begin{equation}
v_{\rm K}= \frac{GM \sqrt{r}}{r-2r_{\rm g}}.
\end{equation}  The excess heat is cooled according to,
\begin{equation}
\Lambda = \frac{f(P-P_{targ})}{\tau_{cool}},
\end{equation} where $f$ is a switch function,$f=0.5[(P-P_{targ})/|P-P_{targ}|+1]$, and $\tau_{cool}$ is the cooling time which we set to the local orbital period.

The maintenance and stability of the target disk scaleheights is of paramount importance in our simulations so that we can clearly isolate any disk height dependences.  Figure \ref{fig-hr_time} shows the time variability in of average scaleheight ratio for the simulations.  For each radial element in each data dump we calculate the geometric scaleheight ratio \begin{equation}
\frac{h(r)}{r} = \Bigg\langle \sqrt{\frac{\int (\theta(r) - \bar{\theta}(r))^2 \rho d\Omega}{\int \rho d\Omega}} \Bigg \rangle,
\end{equation} where $d\Omega=\sin \theta d\theta d\phi$ is the solid angle element in spherical coordinates and, \begin{equation}
\bar{\theta}(r) = \frac{\int \theta(r) \rho d\Omega}{\int \rho d\Omega}
\end{equation} is the average polar angle of the gas.  The total disk-averaged scaleheight is calculated for each instant by weighting every radial bin by its width to account for the nonuniform spacing of the grid in the $r$-coordinate.  This shows the cooling function keeps our target scaleheight throughout the duration of each simulation.  Turbulence introduces small fluctuations in the disk scaleheight ratio, but no worrisome trends are present in the \texttt{hr\_01}, \texttt{hr\_02}, and \texttt{hr\_04} simulations.  In the time trace of the \texttt{hr\_005} simulation there is a residual transient from initialization that artificially inflates the disk, but it decays quickly.  It has no significant influence on our study, but does appear in synthetic light curves presented in Section \ref{sec-obs_sigs} where we discuss it further.

\subsection{Initial Conditions}

The simulations are initialized to a steady-state $\alpha$-disk like solution:
\begin{equation}
\rho(R, \theta)=\rho_{0} R^{-3/2}\exp\Bigg(-\frac{z^2}{2 c_{s}^2 R^3}\Bigg)
\end{equation} where $\rho_0$ is a normalization, $R=r \sin{\theta}$ is the cylindrical radius, and $z= r \cos{\theta}$ is the vertical disk height.  The gas pressure is set from the density, $P = c^2_{\rm s} \rho$.  The velocity field is set such that the azimuthal velocity has the local Keplerian value ($v_{\rm K}$) and $v_r = v_\theta = 0$.   

We set a weak  magnetic field ($\langle \beta \rangle = \langle P_{gas}/P_{mag} \rangle=200$) from a vector potential to guarantee the divergence free condition is satisfied to machine level precision.  The initial field configuration is logarithmically spaced loops with a vertical taper function in the upper atmosphere with the form,
\begin{eqnarray}
&A_r=0, \\
&A_{\theta}=0, \\
&A_{\phi} = A_0 p^{1\over2} e^{-(z/h)^4} R \sin\bigg( \frac{\pi \log(R/2)}{h}\bigg).
\end{eqnarray}  In the early evolution of each model, these field loops feel the radial shear of the disk and the gas is destabilized by the MRI, which goes nonlinear and drives turbulence.

\subsection{Convergence}

To assess the overall convergence and consistency between the models, we apply several resolvability and convergence metrics.  The results of these diagnostics are presented in Table \ref{table-sim_params}.  All averages are taken over the portion of the simulations used the analysis and taken within the volume of the simulation domain contained within two scaleheights above and below the disk where the dynamo behavior originates.

The resolution of the grid in simulation zones per disk scaleheight provides the simplest measure of the resolvability.  Several studies \citep[e.g.][]{2012ApJ...749..189S, 2013ApJ...772..102H} have shown around 30 vertical zones per disk scaleheight offers a crude threshold near where the turbulence is resolved well enough to capture the small-scale evolution, and hence motivated our grid design.  We use the corresponding ``quality factors" to probe how well the simulation grid samples characteristic MRI wavelength,  $\lambda_{MRI}=2\pi v_A /\Omega$, where $v_A$ is the Alfv\'en speed.  For each simulation, we calculate the average quality factor in the $\theta$ and $\phi$ directions, \begin{equation}
Q_\theta=\frac{\lambda_{MRI,\theta}}{R\Delta\theta}
\end{equation}
and \begin{equation}
Q_\phi=\frac{\lambda_{MRI,\phi}}{R\Delta\phi}.
\end{equation}  Values of $Q_\theta>6-8$ have been shown to properly capture the linear growth of the MRI \citep{1995CoPhC..89..127H, 2004ApJ...605..321S, 2010A&A...516A..26F} and a stricter threshold of $Q_\theta>10$ and $Q_\phi>20$ \citep{2011ApJ...738...84H, 2012ApJ...749..189S, 2013ApJ...772..102H} has been established as adequate to capture the nonlinear growth of the instability.  Finally, we measure the saturation of the anisotropic MRI driven turbulence through the average in-plane magnetic tilt angle, \begin{equation}
\Theta_B = -\arctan\bigg(\bigg\langle \frac{B_r}{B_\phi} \bigg\rangle\bigg).
\end{equation}  This quantifies the characteristic orientation of the magnetic field, a value that theoretical estimates predict to be near $\Theta_B\approx15^{\circ}$ \citep{2009ApJ...694.1010G, 2010ApJ...716.1012P}.  Measurements in prior accretion disk simulations \citep[e.g.][]{2011ApJ...738...84H, 2012ApJ...749..189S, 2013ApJ...772..102H, 2016ApJ...826...40H} find the turbulence saturates at a somewhat lower tilt of $\Theta_B\approx11-13^{\circ}$.

By all measures, our models are well-resolved and have similar properties, with the exception of the \texttt{hr\_04} model.  Given the remarkable consistency of the other models, it is difficult to attribute the sudden decrease in the quality factors, effective $\alpha$-parameter, and increase in plasma $\beta$ to the changes in the grid, although we are simulating half of the physical domain (i.e. only $\pm 2.5 \: h/r$).  One clue into the discrepancy is that the inconsistent diagnostics all depend on the strength of the magnetic field, which appears to be lower by roughly half compared to the thermal energy. The $\Theta_B$ value, on the other had, is similar to the other three models, suggesting the saturation of the turbulence is the same, but the field doesn't naturally grow to the same relative level.  A key result of this paper is that the organization of the dynamo in the thicker disks is impeded and less efficient, so the resolution metrics could be biased by this phenomenon.  Nevertheless, when interpreting the following results, it is important to remain aware of what could be a decrease in the effective resolution of the \texttt{hr\_04} model.

\section{Results}
\label{sec-results}

The analysis we conduct is restricted to the final $\Delta t= 3.15\times10^4\:GM/c^3$ ($512$ ISCO orbits) of each simulation.  By allowing the simulations to evolve for at least 100 ISCO orbits, we avoid transient nonphysical behaviors that only exist as the simulated disks relax into a quasisteady state.  The simulation data is written out every $\Delta t = 30.7812\:GM/c^3$, or every half of an orbital period at the ISCO, providing $1024$ snapshots for the analysis.

\subsection{The Global Dynamo}
\label{sec-global_dynamo}

\begin{figure*}
\centering
\includegraphics{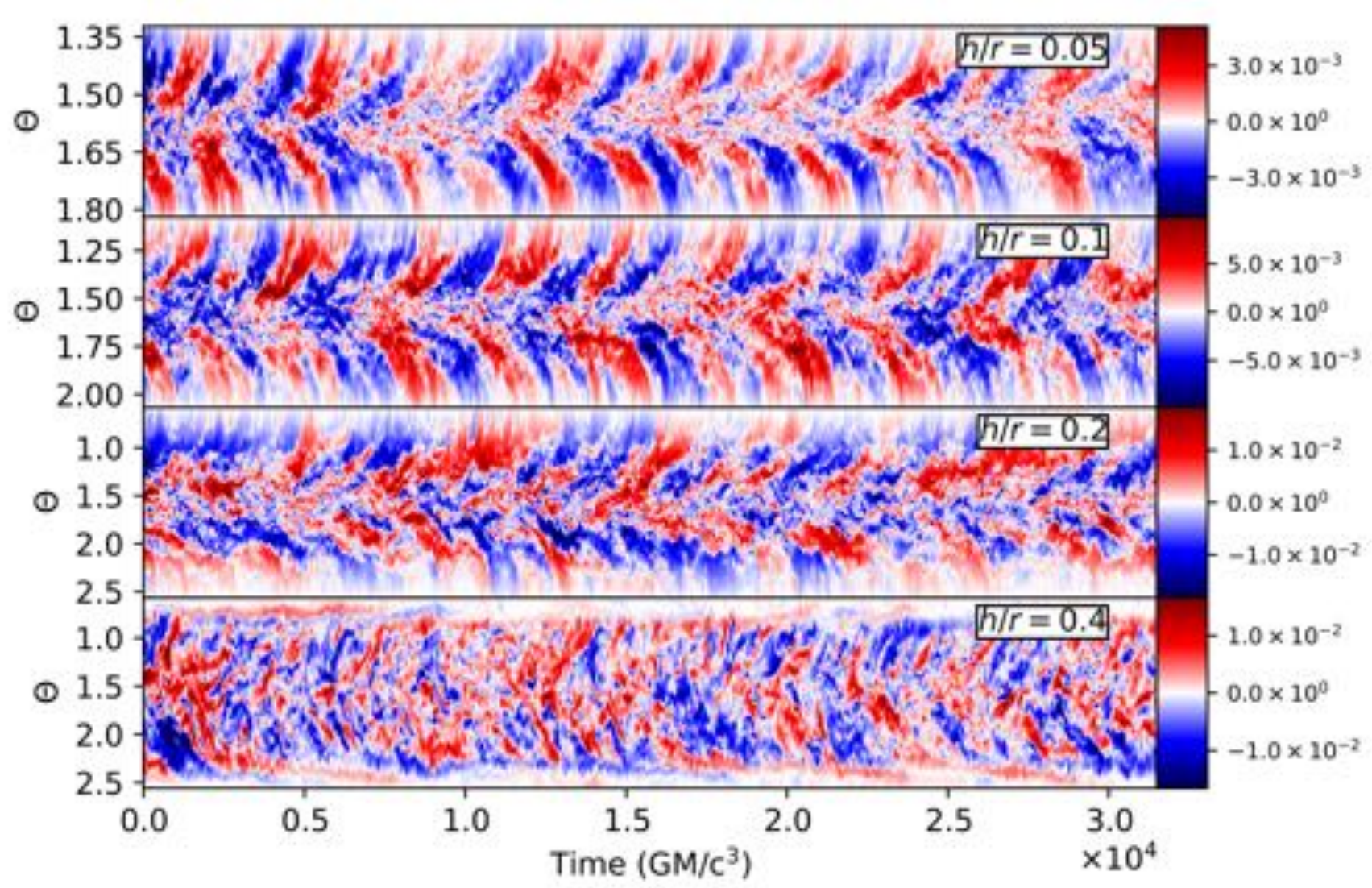}
\caption{Spacetime diagrams of the azimuthally averaged $B_\phi$ at $r=15\:r_g$ for \texttt{hr\_005}(top), \texttt{hr\_01} (second from top), \texttt{hr\_02} (second from bottom), and \texttt{hr\_04} (bottom).  Positive values (red) indicate orientation of the field in the positive $\phi$-direction while negative (blue) indicates an opposite orientation.  Color intensity corresponds to the averaged magnitude.
\label{fig-butterfly_diagrams}}
\end{figure*}

We begin by first presenting spacetime diagrams of the toroidal magnetic field in Figure \ref{fig-butterfly_diagrams}.  The spacetime diagrams were calculated by taking azimuthal averages of $B_\phi$ at each time step at $r=15\:r_g$ in each of our four simulations.  The large-scale dynamo organizes itself into global, vertically stratified sheets of field of similar polarity.  These diagrams effectively trace the evolution of the cross section of this pattern at a chosen radii, which typically reveals a vertically propagating pattern that has a characteristic acceleration, evidenced by the increasing slope.  Snapshots of the normal pattern's structure are presented in Figure 9 of \citet{2016ApJ...826...40H}.

The different behaviors between the models are immediately apparent with the most striking difference being the breakdown of the regular, periodic oscillation pattern with increasing disk thickness.  The thinnest disk in the \texttt{hr\_005} model shows a cyclical building of the field and decay with a reversal of the orientation.  Near the midplane it is fairly stochastic, but in the atmosphere the organization develops as the overmagnetized regions buoyantly rise.  In the \texttt{hr\_01}  model, a pattern is still present, but it is less well organized compared to the \texttt{hr\_005} simulation.  In \texttt{hr\_02} and \texttt{hr\_04} there are hints the organized oscillations attempt to develop, but patchy, chaotic fluctuations dominate the behavior.  However, there is still field amplification and regeneration throughout the simulation, despite the lack of a regular, ordered pattern.  

\begin{figure*}
\centering
\includegraphics{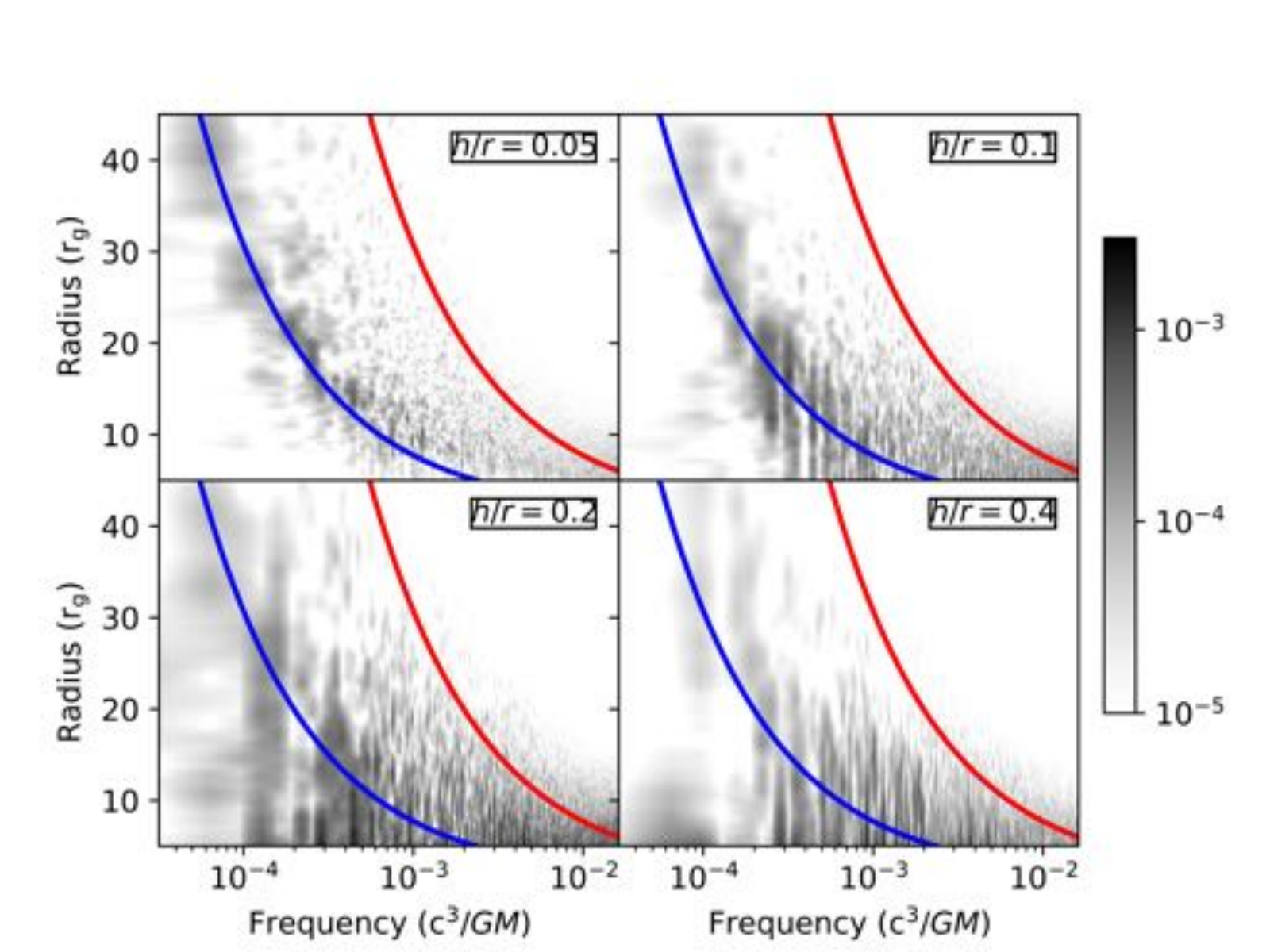}
\caption{PSDs of $B_{\phi}$ at $1.5h$ above the disk midplane for \texttt{hr\_005} (top left), \texttt{hr\_01} (top right), \texttt{hr\_02} (bottom left), and \texttt{hr\_04} (bottom right).  Darker colors (black) represents greater power in that frequency bin for a given radius.  The orbital frequency is shown with the red line and  ten times the orbital frequency is shown with the blue line.
\label{fig-2D_FFT}}
\end{figure*}

Further inspection of the \texttt{hr\_005} and \texttt{hr\_01} models reveal other interesting features of the butterfly pattern.  There are periods of the simulation where the dynamo seemingly fails to reverse, for instance $t=1.5\times10^4\:GM/c^3$ and $t=1.7\times10^4\:GM/c^3$ in  \texttt{hr\_005}.  Additionally, the large-scale field generated by the dynamo can evolve independently in the upper and lower coronal regions of each disk.  At some points in the simulations the magnetic fields are aligned in the upper atmospheres, like at $t=6.0\times10^3 GM/c^3$ in the \texttt{hr\_005} simulation, but then at a later time the fields are antialigned, like $t=2.1\times10^4 GM/c^3$.  This is the same change in parity observed in \citet{2012ApJ...744..144F}.  Several factors contribute to this, including irregularities in the local evolution of the dynamo cycle and a global influence from the coupling of field since the field buoyantly rises as coronal sheets of field with similar polarity \citep{2011MNRAS.416..361B, 2016ApJ...826...40H}.  These butterfly diagrams can be compared to those observed from local shearing box simulations, e.g. \citet{2010ApJ...713...52D} and \citet{2012MNRAS.422.2685S}, which tend to have more more regular periods.

\begin{deluxetable*}{ccccc}
\tabletypesize{\small} 
\vspace{-0.5cm}
\tablewidth{0pt} 
\tablecaption{ Dynamo Coefficient Fits \label{table-alpha_dyn_fit}} 
\tablehead{ 
\colhead{Simulation} & \colhead{$\alpha_{d, uh}$} & \colhead{Offset$_{uh}$} & \colhead{$\alpha_{d, lh}$} & \colhead{Offset$_{lh}$}}
\startdata 
\texttt{hr\_005} & $(-9.5\pm0.1)\times10^{-5}$ & $(-0.4\pm1.2)\times10^{-8}$ & $(1.0\pm0.1)\times10^{-4}$ & $(0.4\pm1.5)\times10^{-8}$\\
\texttt{hr\_01} & $(-2.1\pm0.2)\times10^{-4}$ & $(-1\pm7)\times10^{-8}$ & $(1.7\pm0.2)\times10^{-4}$ & $(-1\pm7)\times10^{-8}$\\
\texttt{hr\_02} & $(-2.2\pm0.4)\times10^{-4}$ & $(-4\pm2)\times10^{-7}$ & $(2.2\pm0.4)\times10^{-4}$ & $(-1.9\pm1.8)\times10^{-7}$ \\
\texttt{hr\_04} & $(-2.3\pm0.5)\times10^{-4}$ & $(-2.1\pm1.0)\times10^{-7}$ & $(1.7\pm0.4)\times10^{-4}$ & $(-1.5\pm1.2)\times10^{-7}$
\enddata 
\end{deluxetable*}

Turning to the thicker disks in the \texttt{hr\_02} and \texttt{hr\_04} models, we see the amplification of the field is typically localized.  Enhanced regions of strong field form, but they are disrupted before they can collect in the midplane.  In both of these simulations we see that even though the pockets of strengthened field do not trace out a butterfly pattern \emph{per se}, they still typically originate near the midplane and are expelled into the disk atmosphere, presumably due to their magnetic buoyancy like before.  In the \texttt{hr\_02} model, there are periods when the butterfly pattern almost takes hold, but it often only in one hemisphere and traces out an inconsistent rise.

The power spectral density (PSD) at one scaleheight above the disk midpane, shown in Figure \ref{fig-2D_FFT}, more clearly show the presence or absence of periodicity in the dynamo.  The PSDs were calculated as $P(\nu)=\nu |\widetilde{f}(\nu)|^{2}$ where $\widetilde{f}(\nu)$ is the Fourier transform of the time sequence of the variable of interest,
\begin{equation}
\widetilde{f}(\nu) = \int f(t) e^{-2 \pi i \nu t} dt.
\end{equation}  Here, we consider the power spectra of the azimuthally averaged $B_\phi$ at each radial grid point.  We also average $B_\phi$ over the $\theta$ direction from $\pi/2-1.25\:h/r$ to $\pi/2-0.75\:h/r$ to get a better sense of the dominating mean field.

The \texttt{hr\_005} and \texttt{hr\_01} simulations both show a distinct band of power at one tenth of the orbital frequency similar to the dynamo periods found in \citet{2010MNRAS.405...41G}, \citet{2010ApJ...713...52D}, \citet{2011MNRAS.416..361B}, \citet{2011ApJ...730...94S}, and \citet{2012MNRAS.422.2685S}.  This band of enhanced power spans roughly a factor of three in frequency space and extends radially approximately $10\:r_g$, consistent with the PSDs seen in thin accretion simulations, like those presented in \citet{2011ApJ...736..107O} and \citet{2016ApJ...826...40H}.  At lower frequencies there is very little power and at higher frequencies there is residual power up to a dissipative scale.

The power spectra of the \texttt{hr\_02} and \texttt{hr\_04} simulations tell a different story, though.  In these simulations there are no discernible bands of power indicating they lack any unique timescales where the field has significant oscillations.  There is power at all frequencies below the orbital frequency, indicating the large-scale dynamo is operating; however, it is neither organized nor confined to a specific timescale.  This confirms the seeming randomness and disorder seen in the respective butterfly diagrams, suggesting there is no single characteristic scale on which energy is injected, rather the disorder allows the flow to injecting energy over a range of scales, which then decay.  Nevertheless, the amplitude of the power is nearly equal between all of the models, indicating that there is not a dearth of power in the low-frequency field fluctuations in the thicker disk simulations.

\subsection{Measuring $\alpha_{\rm d}$}

Next, we seek to probe the heart of the dynamo mechanism by measuring the parameterization of the ``$\alpha$-effect."  The large-scale dynamo is typically interpreted through a mean field theory.  To produce the large scale toroidal field, there must be a net electromagnetic force (EMF) to induce a magnetic field in the azimuthal direction, which is predominately governed by:
\begin{equation}
\langle \mathcal{E'}_\phi \rangle = \alpha_{d,\phi\phi} \langle B_\phi \rangle.
\end{equation}  Like other works (e.g. the local shearing box simulations of \citet{1995ApJ...446..741B}, \citet{1997MNRAS.288L..29B}, \citet{2001A&A...378..668Z}, and \citet{2010ApJ...713...52D} and global disk model of \citet{2012ApJ...744..144F}), we neglect contributions to the $\alpha$ effect from the less influential $\alpha_{d,rr}$ component and also higher order derivatives of the magnetic field from the diffusivity tensor (usually denoted as $\tilde{\eta}$), thereby allowing us to approximate $\alpha_{\phi\phi}$ through a simple correlation between the turbulent EMF,
\begin{equation}
\mathcal{E'_{\phi}}=v_r' B_\theta' - v_\theta' B_r',
\end{equation} where $X'$ of a flow variable $X$ indicates its fluctuating component taken by subtracting off its azimuthal average ($X'=X-\langle X \rangle$), and the average toroidal magnetic field.

\begin{figure*}
 \vspace{-0.2cm}
\centering
  \subfigure[Upper Hemisphere \texttt{hr\_005}]{\includegraphics[width=0.9\textwidth]{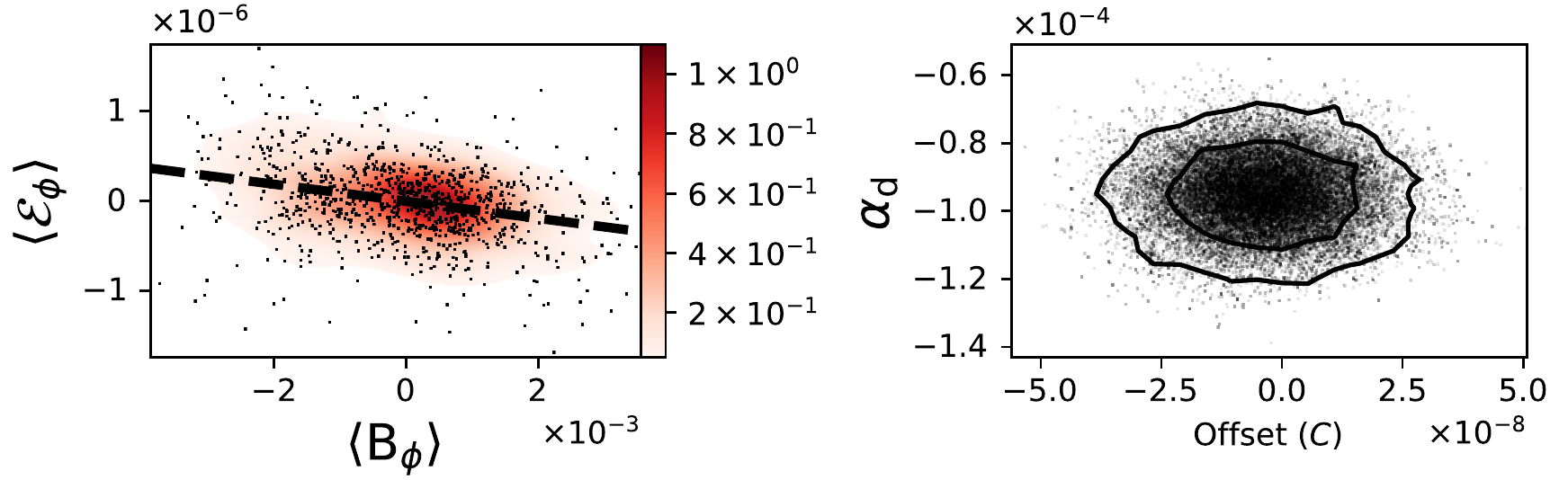}}
    \centering
  \subfigure[Upper Hemisphere \texttt{hr\_01}]{\includegraphics[width=0.9\textwidth]{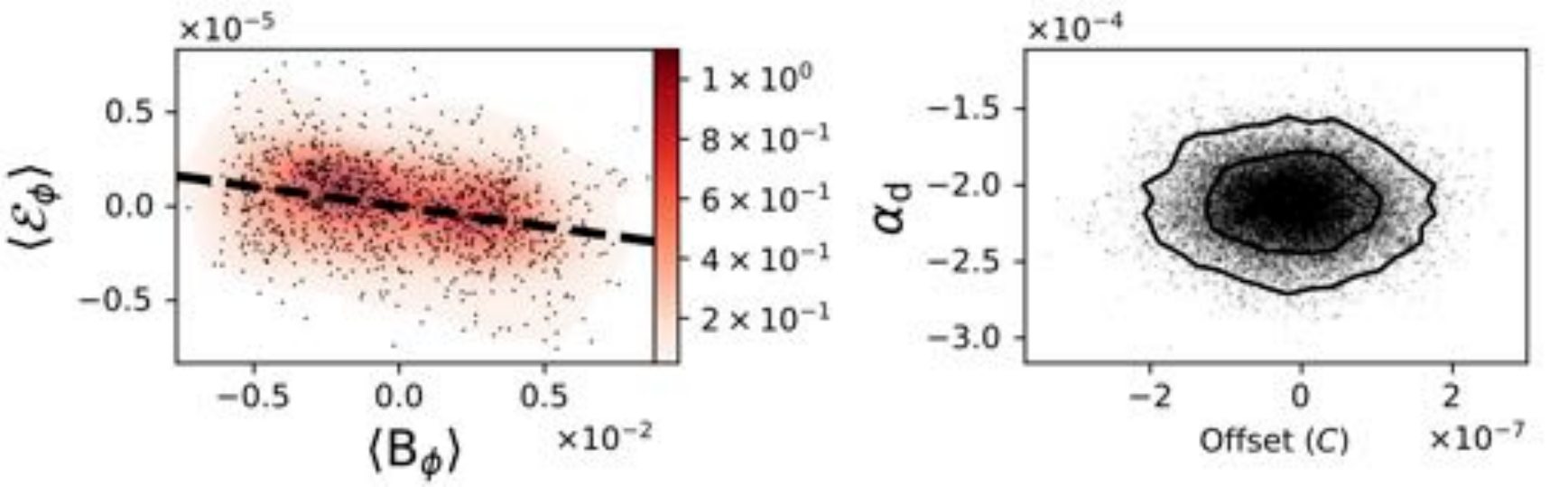}}
    \centering
  \subfigure[Upper Hemisphere \texttt{hr\_02}]{\includegraphics[width=0.9\textwidth]{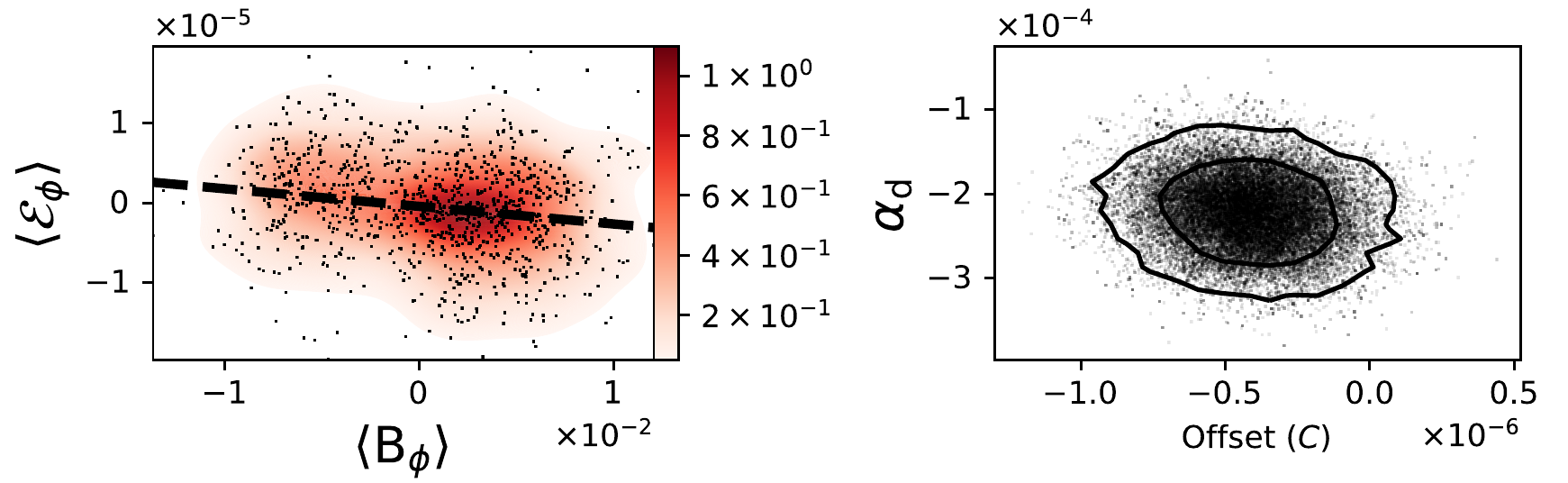}}
    \centering
  \subfigure[Upper Hemisphere \texttt{hr\_04}]{\includegraphics[width=0.9\textwidth]{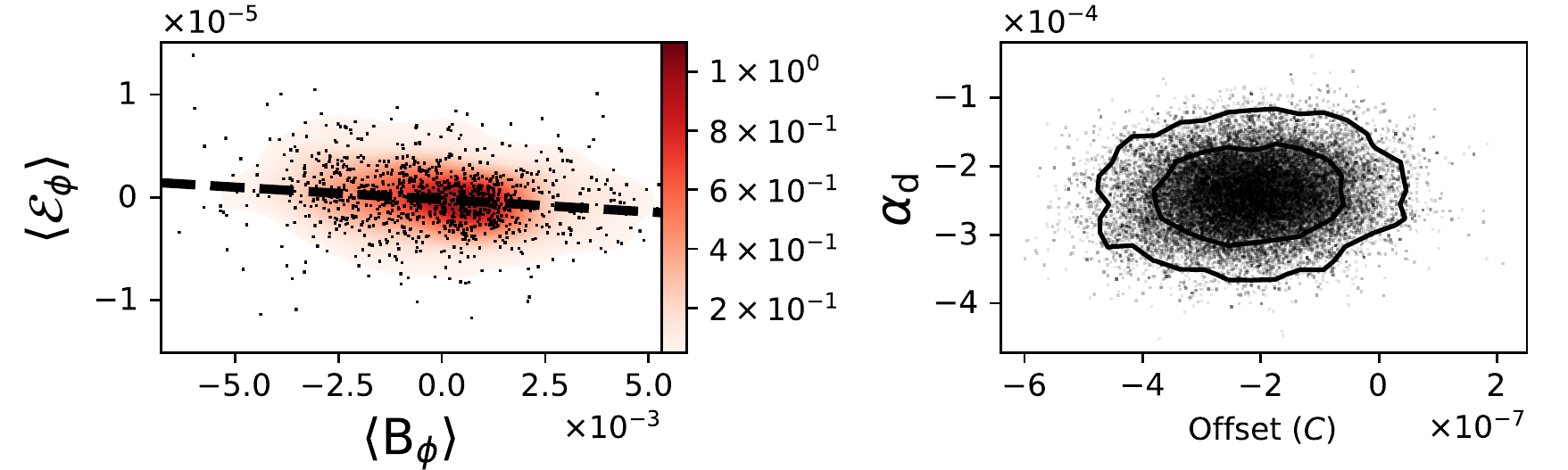}}
\caption{Scatter plots of instantaneous values of $\langle B_\phi \rangle$ vs $\langle \mathcal{E'}_\phi \rangle$ in the upper coronal regions of the disks (black dots) with the best fitted lines (left column) and the best parameter fits from our MCMC modeling with $1\sigma$ and $2\sigma$ contours (right column).  The color coding in the lefthand panels shows the density distribution of the points, estimated from a Gaussian kernel.
\label{fig-upper_hem_fit}}
\end{figure*}

\begin{figure*}
 \vspace{-0.2cm}
 \centering
  \subfigure[Lower Hemisphere \texttt{hr\_005}]{\includegraphics[width=0.9\textwidth]{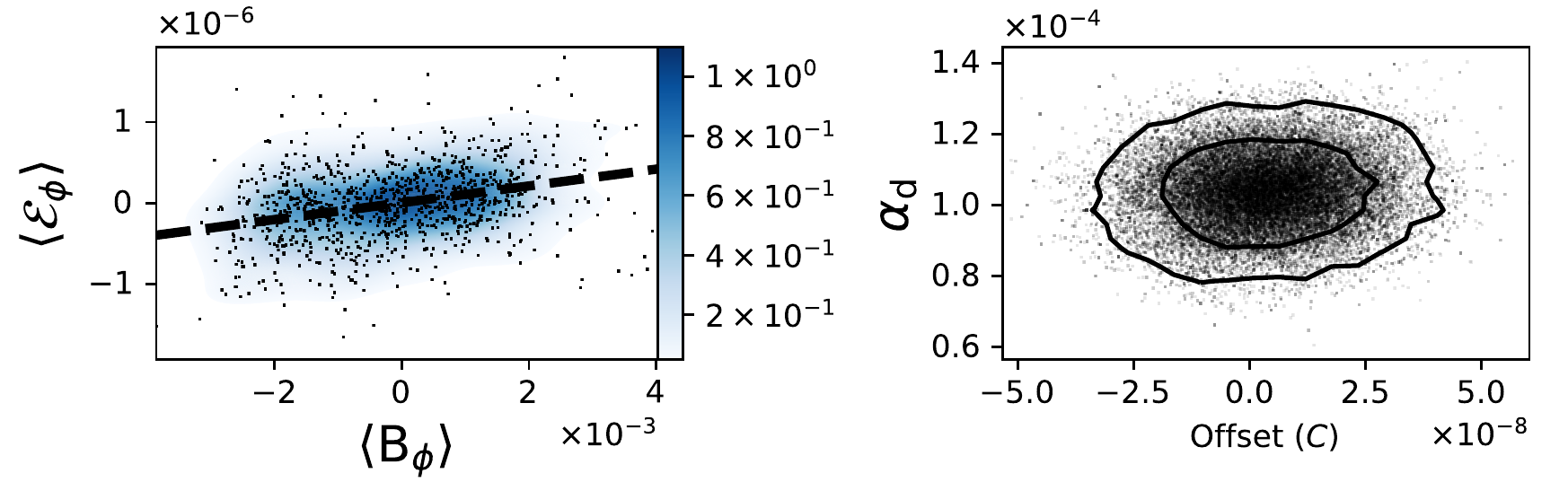}}
    \centering
  \subfigure[Lower Hemisphere \texttt{hr\_01}]{\includegraphics[width=0.9\textwidth]{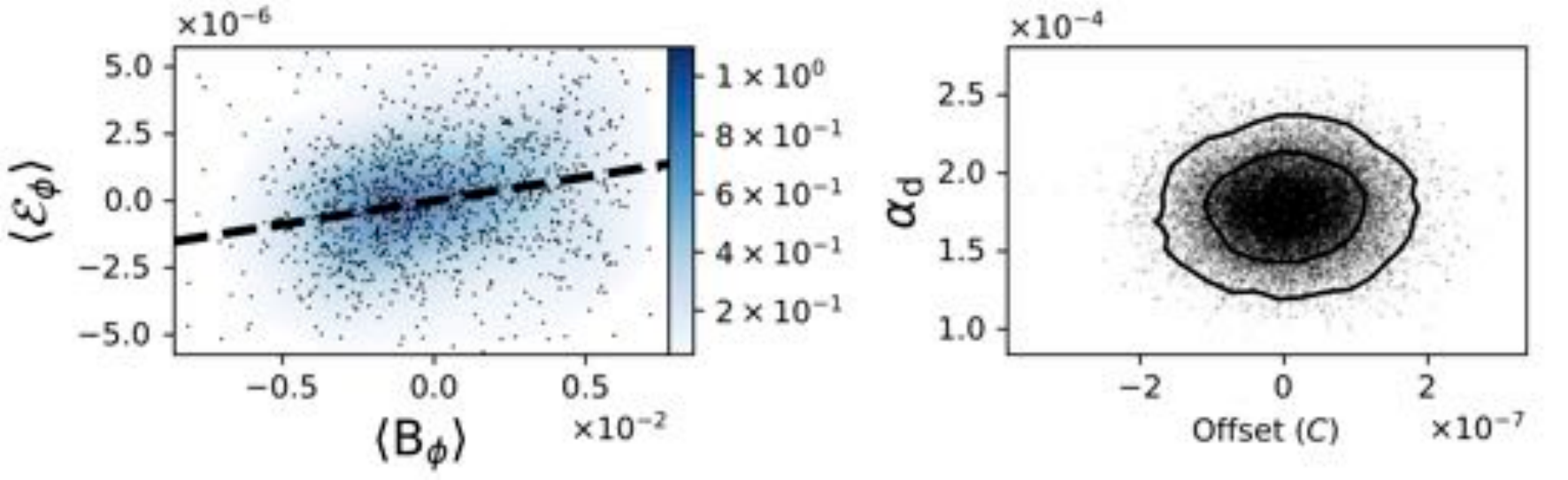}}
    \centering
  \subfigure[Lower Hemisphere \texttt{hr\_02}]{\includegraphics[width=0.9\textwidth]{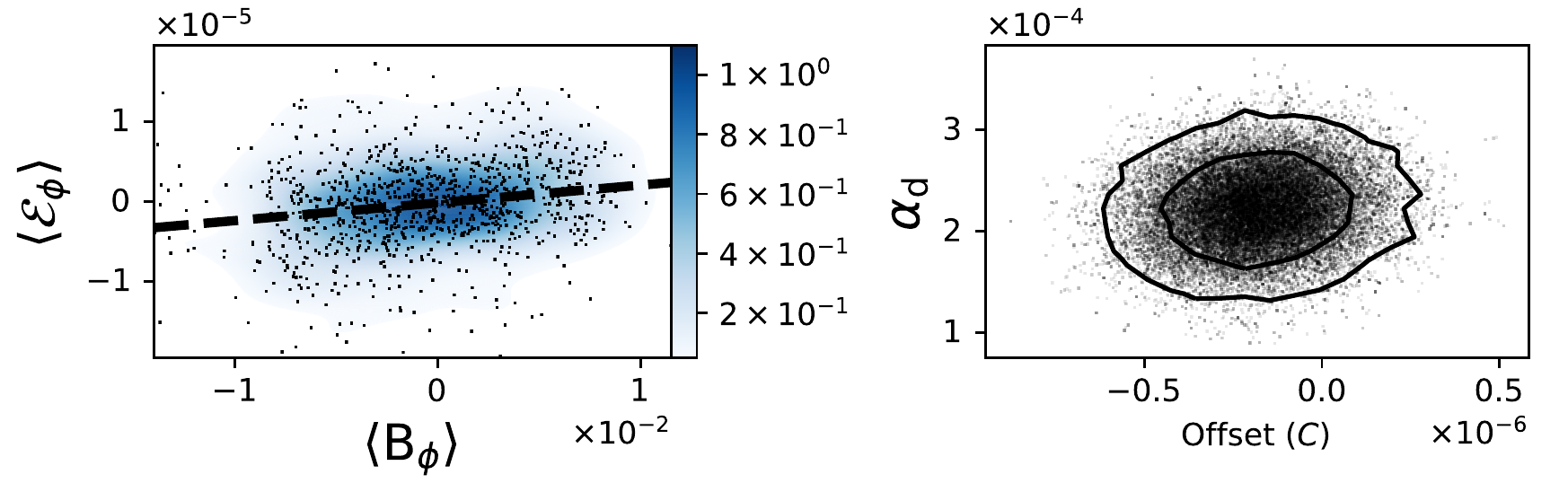}}
    \centering
  \subfigure[Lower Hemisphere \texttt{hr\_04}]{\includegraphics[width=0.9\textwidth]{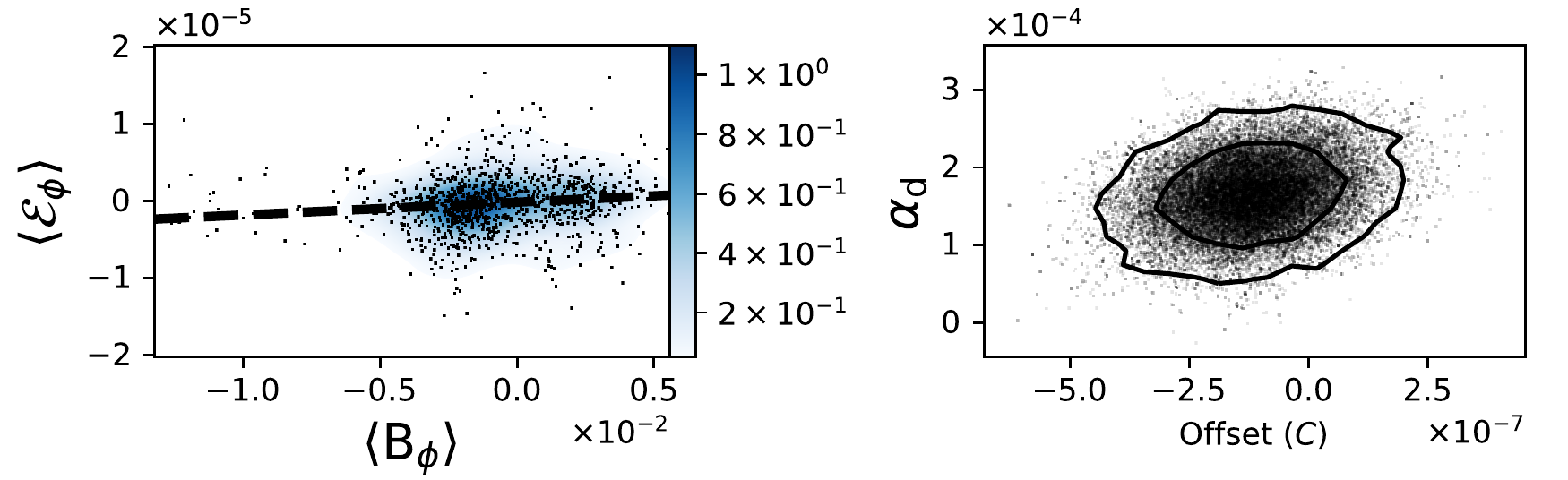}}
\caption{Scatter plots of instantaneous values of $\langle B_\phi \rangle$ vs $\langle \mathcal{E'}_\phi \rangle$ in the lower coronal regions of the disks (black dots) with the best fitted lines (left column) and the best parameter fits from our MCMC modeling with $1\sigma$ and $2\sigma$ contours (right column).  The color coding in the lefthand panels shows the density distribution of the points, estimated from a Gaussian kernel.
\label{fig-lower_hem_fit}}
\end{figure*}

We calculate the volume averaged $\mathcal{E'_{\phi}}$ and $B_\phi$ in the upper and lower hemispheres of each simulation for each data dump.  The regions in which we take these averages spans the entire azimuthal domain, radially from $r=15\:r_g$ to $r=15+15(h/r)$, and poloidally from $h$ to $2h$ in the upper hemisphere and $-h$ to $-2h$ in the lower hemisphere.  The correlation of $\langle \mathcal{E'_{\phi}} \rangle$ and $\langle B_\phi \rangle$ is then measured by fitting a simple line of the form,
\begin{equation}
\langle \mathcal{E'}_\phi \rangle = \alpha_{d} \langle B_\phi \rangle + C,
\end{equation} where $C$ is an allowed vertical offset.  The fitting was done using a Bayesian MCMC method.  Figure \ref{fig-upper_hem_fit} shows the data, best fit, and parameter contour plots of the fit parameters for the upper hemisphere of the different models, and Figure \ref{fig-lower_hem_fit} shows the corresponding results for the lower hemispheres.  Table \ref{table-alpha_dyn_fit} summarizes the best fits.

We find that $\alpha_d$ consistently has an amplitude of $|\alpha_d|\approx1-2\times10^{-4}$ in all of these disk simulations with the upper hemisphere having a negative sign and the lower hemisphere having a positive sign, similar to \citet{1995ApJ...446..741B} and \citet{2010ApJ...713...52D}, but opposite of other works.  Furthermore, they are a slightly weaker than other values reported in the literature.  The increase in the scatter of the $\langle \mathcal{E'_{\phi}} \rangle$ and $\langle B_\phi \rangle$ correlations with increased disk thickness indicates the diffusive term becomes a larger contributor.

\subsection{Magnetic Helicity}

\begin{figure*}
  \centering
\includegraphics{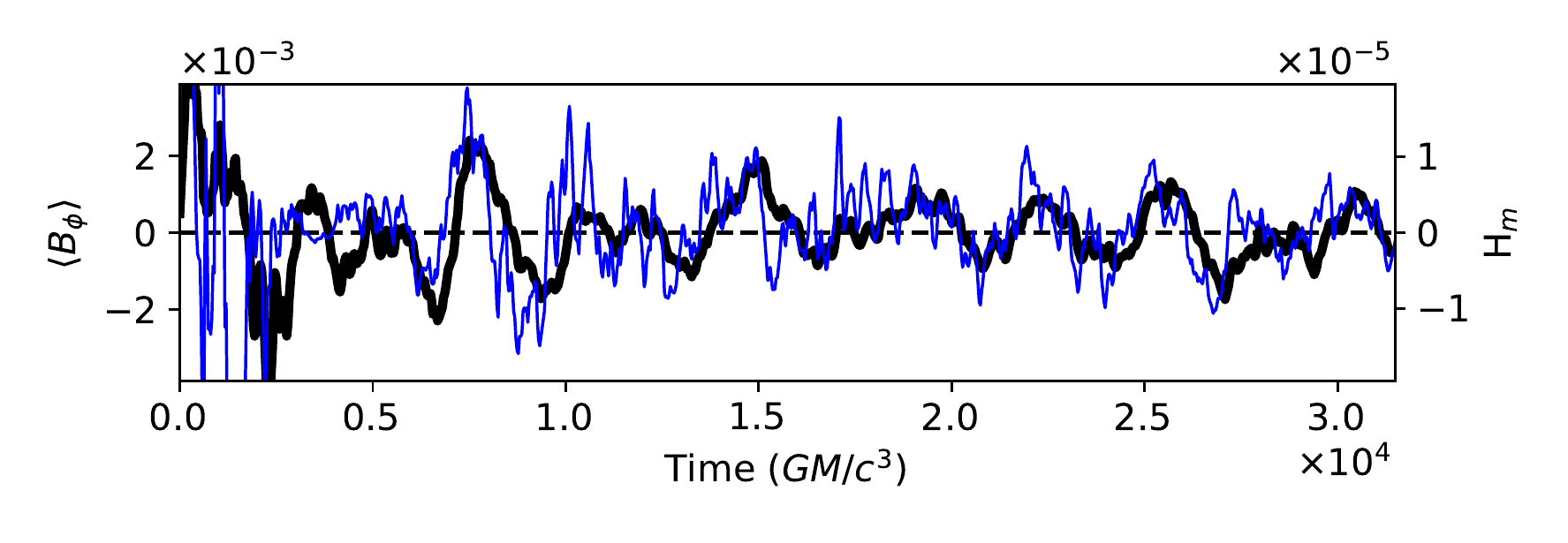}
\caption{Average $B_\phi$ (dark black line) and $\mathcal{H}_m$ (thin blue line) for \texttt{hr\_005} at $r=15\:r_g$.  The two time traces have a Pearson-r statistic of $0.4$.
\label{fig-helicity}}
\end{figure*}

As an exploratory exercise to bore into the organization, or lack thereof, in the dynamo pattern, we sought to measure the large- and small-scale magnetic helicities, current helicities, and kinetic helicities in the four simulations.  Teasing out the time variability of these quantities from the chaotic turbulence ultimately proved not to be feasible with these models, and we were not able to overcome the inherent noise from differencing the stochastically fluctuating fluid variables.  However, we were able to find a correlation between the large-scale magnetic helicity density,\begin{equation}
\mathcal{H}_m= \langle \bar{\bf{A}}\cdot\bar{\bf{B}}\rangle
\end{equation} with the average $\langle B_\phi \rangle$ in the midplane of the \texttt{hr\_005} simulation, shown in Figure \ref{fig-helicity}.  Since the large-scale helicity density is a volume integrated quantity, it was calculated in a subdomain of the global simulation.  We chose a reference radius, $r_{ref}=15\:r_g$, and then defined the subdomain to span radially from $r_{ref}$ to $r_{ref}(1+h/r)$, vertically from the midplane ($\theta=\pi/2$) to one disk scaleheight ($\theta=\pi/2-h/r$), and in azimuth over the entire $\phi$-domain.  The subdomain was designed this way for several reasons.  The radial location was selected to be far enough from the ISCO that turbulent edge effects \citep{2002ApJ...573..754K} are miniscule.  Moving away from the inner boundary also has the additional benefit that the dynamo period is longer, so our effective time resolution is increased by approximately a factor of 4, but it is still short enough we can still study the behavior for many evolutionary times.  We are primarily interested in where the dynamo pattern originates and first organizes itself, which drives us to the midplane of the disk.  However, since the helicity production is roughly asymmetric about the midplane \citep{2012PhyS...86e8202B}, we must only sample one hemisphere so that there is not cancelation, and we choose the upper hemisphere by default.  We tested several vertical extents and determined that the general trends held no matter where we placed the upper vertical boundary of the subdomain within one disk scaleheight above the midplane, but the noise was minimized at the upper limit.

Typically, $\mathcal{H}_m$ evolves with $\langle B_\phi \rangle$.  When $\langle B_\phi \rangle$ oscillates and changes sign, $\mathcal{H}_m$ seems to vary with it.  There are periods when they cycle seems to stall, like from $t=1.0\times10^4\:GM/c^3$ to $t=1.3\times10^4\:GM/c^3$, where $\mathcal{H}_m$ similarly shows no real evolution.  This is either because the helicity is not produced or because there is no \emph{net} production due global effects like cancelation with neighboring radii.  Noise dominates much of the signal, but the linear correlation of $\mathcal{H}_m$ and $\langle B_\phi \rangle$, as measured with the Pearson-r statistic, is $r=0.38$ which indicated a moderate correlation.  Over different periods of the simulation it changes, though.  For instance, if the correlation coefficient is only calculated over the last half of the simulation, it is higher at $r=0.58$.

\begin{figure*}
  \centering
  \subfigure[\texttt{hr\_005} Mass Accretion Rate Distribution]{\includegraphics[width=0.45\textwidth]{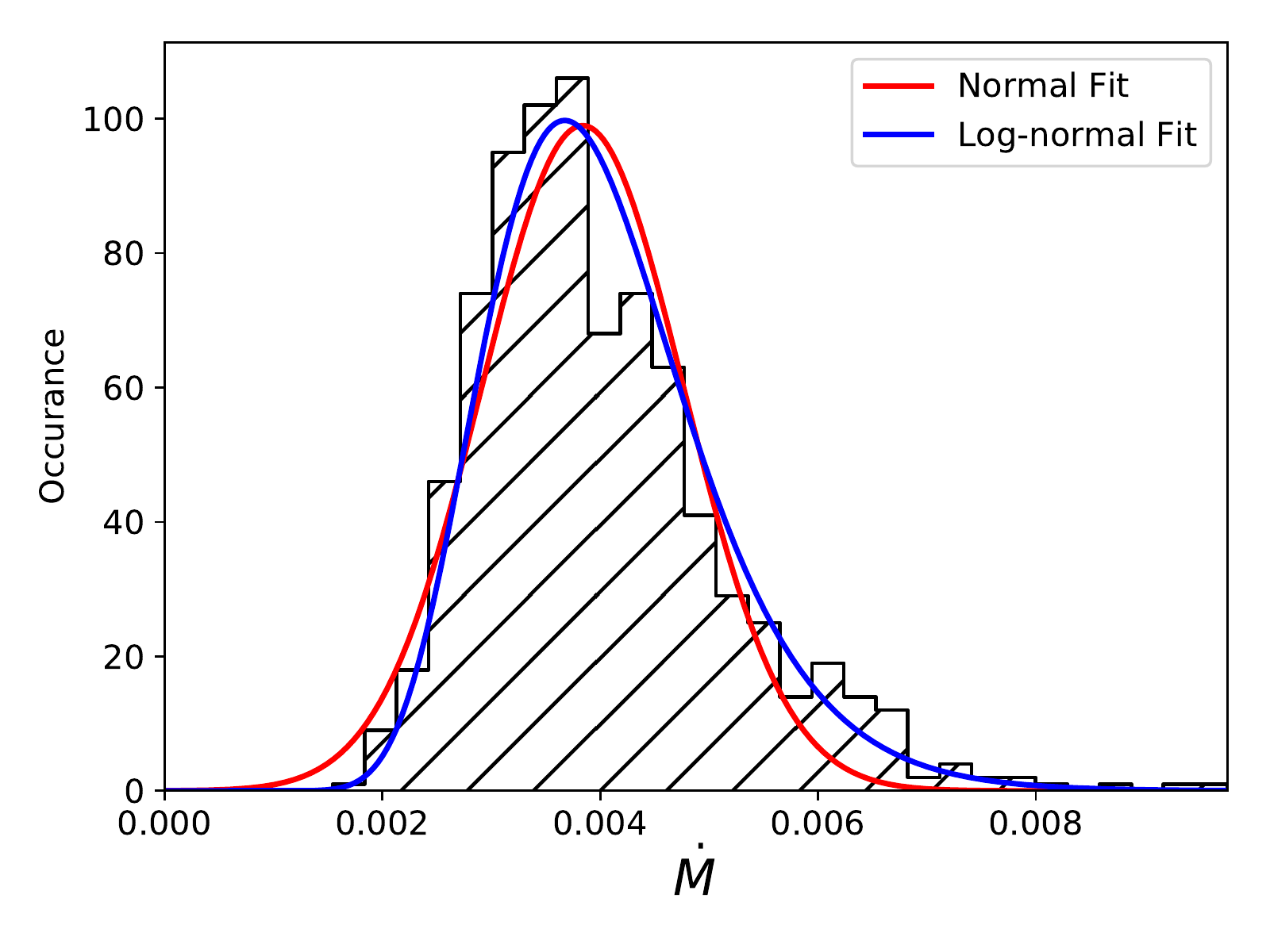}}
    \centering
  \subfigure[\texttt{hr\_01} Mass Accretion Rate Distribution]{\includegraphics[width=0.45\textwidth]{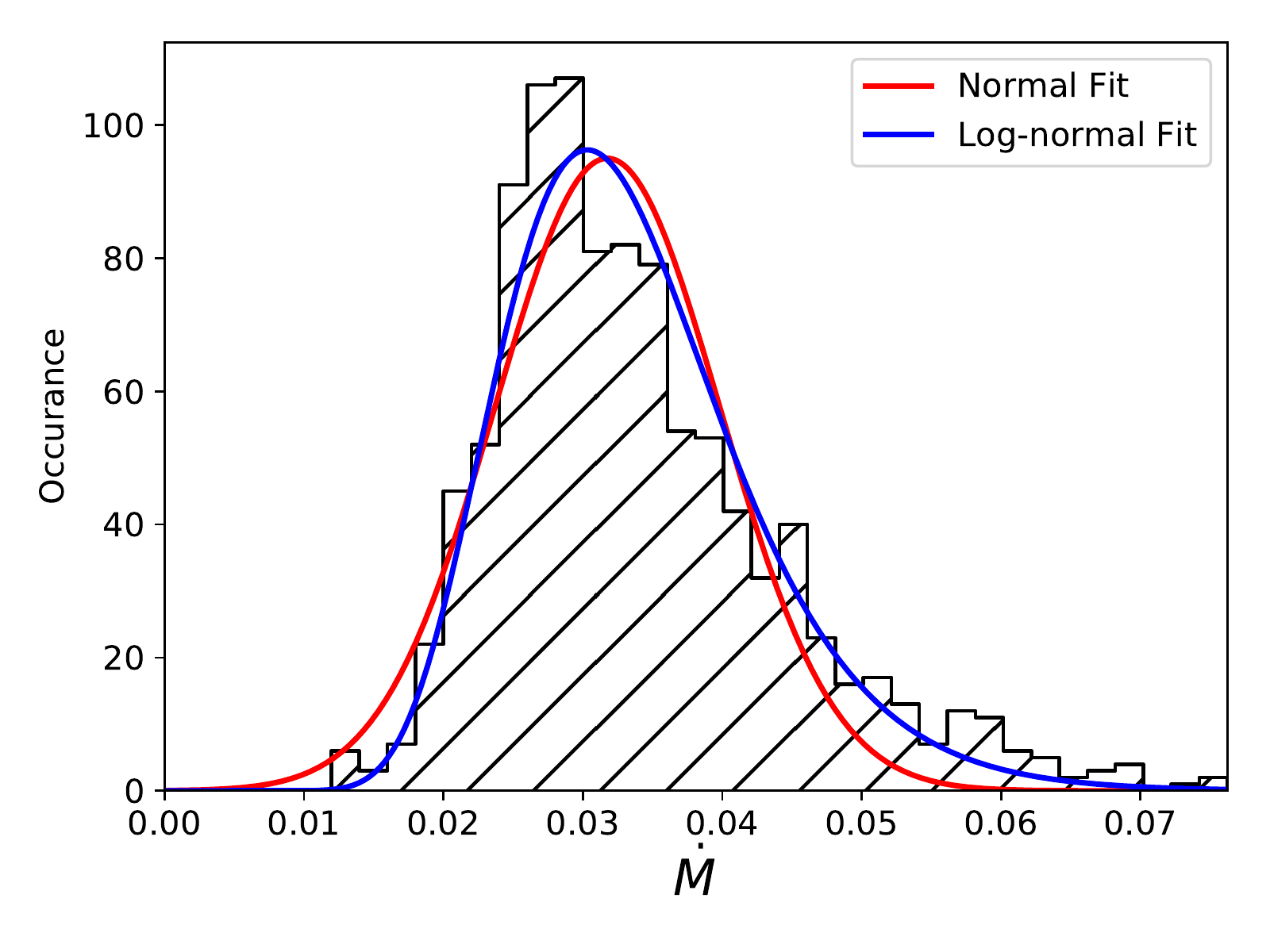}}
    \centering
  \subfigure[\texttt{hr\_02} Mass Accretion Rate Distribution]{\includegraphics[width=0.45\textwidth]{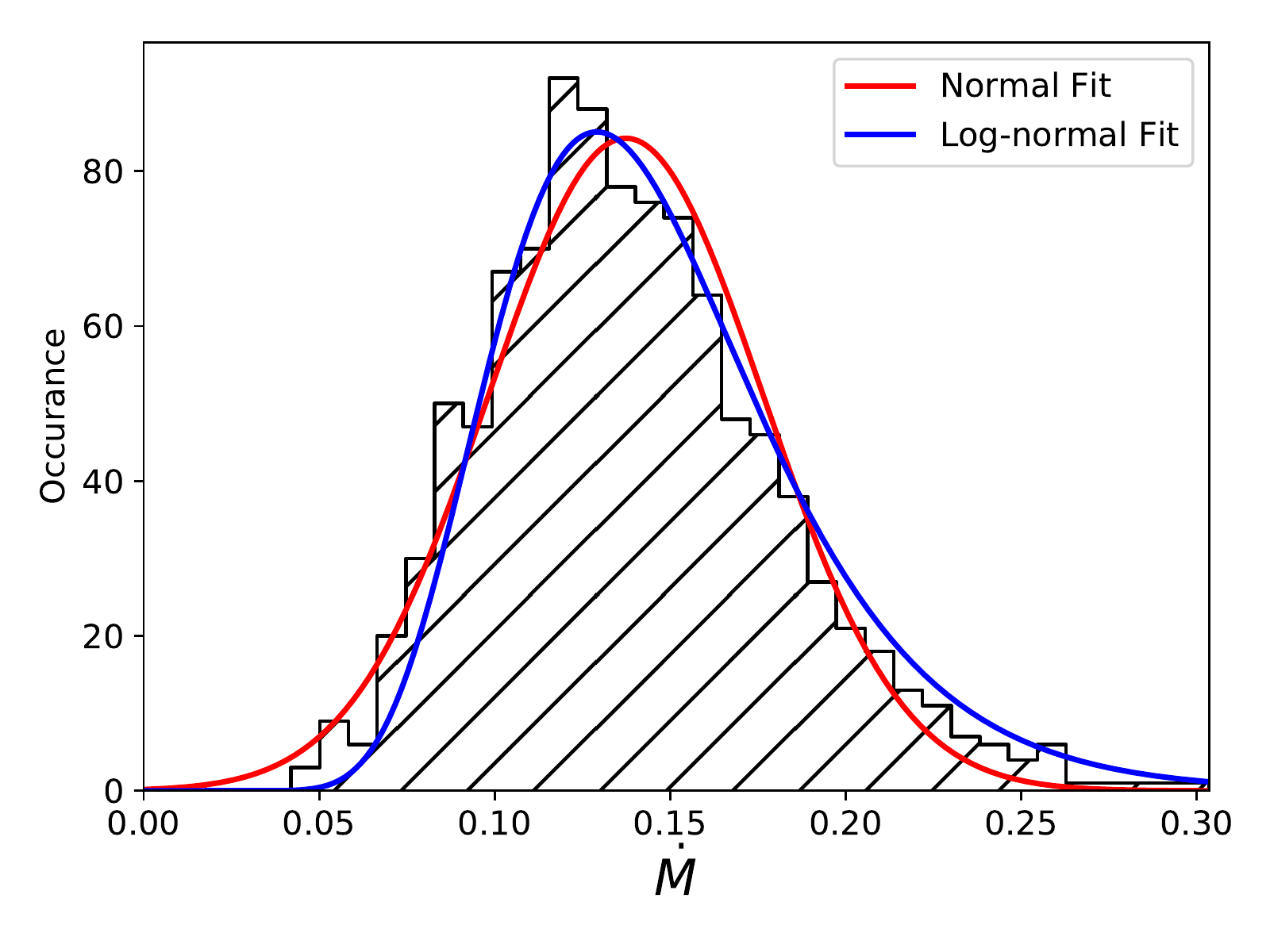}}
    \centering
  \subfigure[\texttt{hr\_04} Mass Accretion Rate Distribution]{\includegraphics[width=0.45\textwidth]{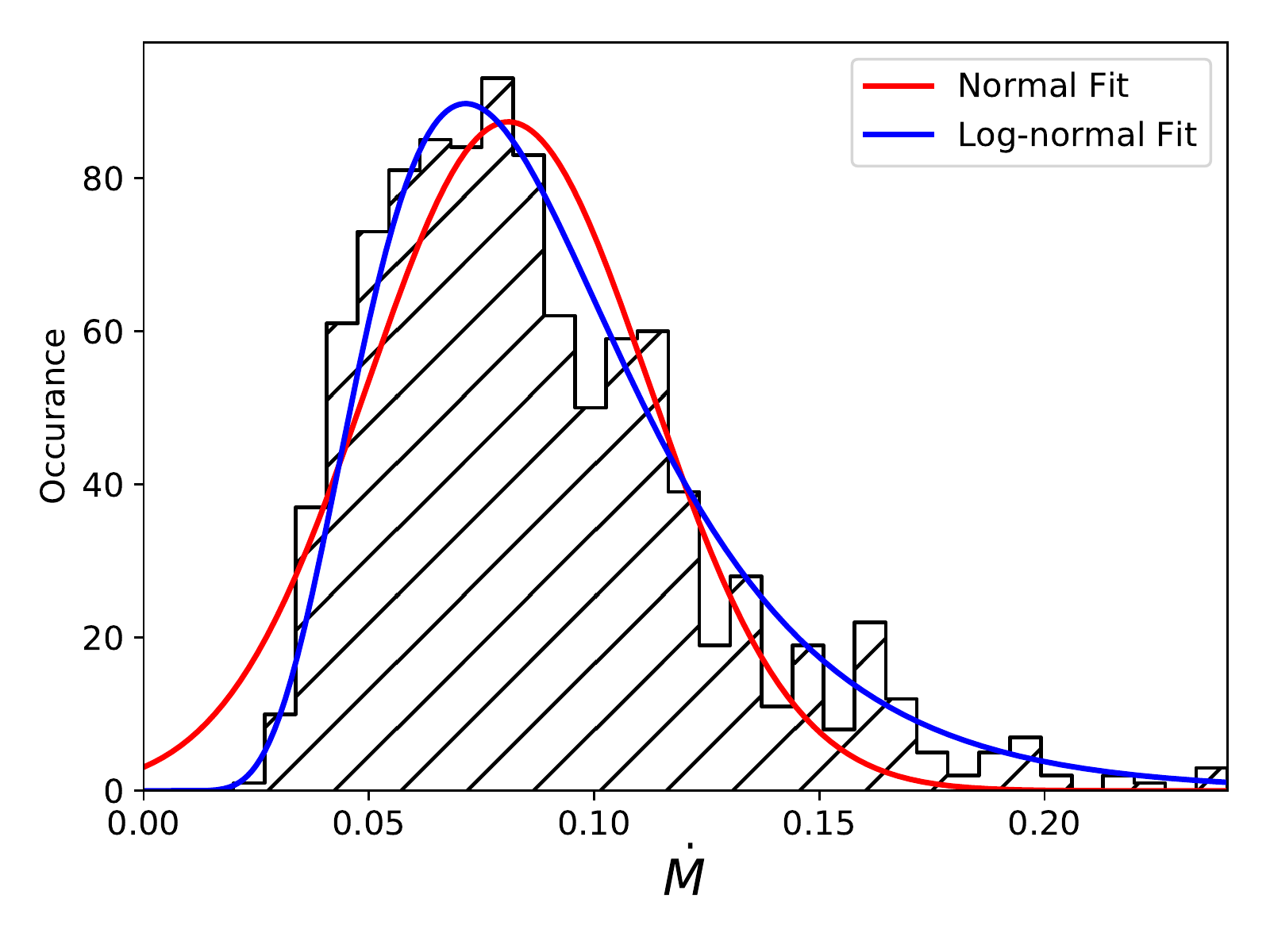}}
\caption{Mass accretion rate distributions at the inner boundary of \texttt{hr\_005} (a), \texttt{hr\_01} (b), \texttt{hr\_02} (c), and \texttt{hr\_04} (d).  The distributions have been fit by a Gaussian function (red) and log-normal function (blue lines), and they are all better fit by a log-normal function.  As noted in the text, we are primarily concerned with the qualitative shape rather than the quantitative amplitude since all simulations were initialized with the same density normalizations and the total disk ``mass" scales with increasing thickness.
\label{fig-mdot_fits}}
\end{figure*}

This hint that the large-scale magnetic helicity is tied to the organization of the dynamo should help motivate further study of the role that the flow helicity plays in regulating the global dynamo in future investigations.  There is a wealth of literature showing a connection between the dynamo and its quenching to the flow helicity from analytic studies \citep{2002ApJ...579..359B, 2005PhR...417....1B, 2001ApJ...550..752V}.  Explaining the large-scale accretion disk dynamo through this lens offers great prospect and could yield a greater understanding of the global disk evolution.  Of particular interest is how the helicity produced at different radii, and consequently on different timescales, interacts.  Unlike the magnetic field which decays through turbulence to larger wave numbers, the magnetic helicity undergoes an inverse cascade and relaxes to smaller wave numbers, which could effectively couple radii as the helicities add or cancel, depending on the interference of the production patterns.  Unraveling these connections may explain some of the radial coherence, as well as the intermittencies and irregularities observed in global dynamos.

\subsection{Mass Accretion Rates}

Having seen the role of organization on the magnetic field, we can begin to ask how it effects the disk evolution.  One of the clearest ways to detect the dynamo influence is in the mass accretion rate \begin{equation}
\dot{M} = \int \rho v_{R} R \textrm{sin}(\theta) d\phi d\theta,
\end{equation} where we take $R$ to be the inner simulation boundary, calculated from the instantaneous values of density and velocity for each data dump.  In \citet{2016ApJ...826...40H} we showed that the large-scale dynamo drove propagating fluctuations which appear as a telling log-normal distribution in the large-scale accretion rate due to the modulation of the effective disk viscosity.  Figure \ref{fig-mdot_fits} shows the histograms of $\dot{M}$ for each of the models.  Note, that in the \texttt{hr\_005} model, there is a residual transient in $\dot{M}$ from the initialization so the first $100$ orbits of the analysis which has been removed to prevent contamination of the histogram.  In all four of our models the histograms have a characteristic skewed distribution.  We fit both normal 
\begin{equation}
P(x)_N = P_0e^{{{ - \left( {x - \mu } \right)^2 } \mathord{\left/ {\vphantom {{ - \left( {x - \mu } \right)^2 } {2\sigma ^2 }}} \right. \kern-\nulldelimiterspace} {2\sigma ^2 }}}
\end{equation} and log-normal,
\begin{equation}
P(x)_{LN} = \frac{P_0}{x}e^{{{ - \left( {lnx - \mu } \right)^2 } \mathord{\left/ {\vphantom {{ - \left( {x - \mu } \right)^2 } {2\sigma ^2 }}} \right. \kern-\nulldelimiterspace} {2\sigma ^2 }}}
\end{equation} functions to the $\dot{M}$ distributions, where $x$ is the data count rate in each bin, $\sigma$ is the distribution width, and $\mu$ is the peak.  In every case they are better fit by the fast rise and slow decay in the high valued tail of the log-normal distribution.  The normal distributions fail to match the shapes of the distributions at high and low $\dot{M}$ and consistently are shifted to the right of the peak in the distribution.  As a reminder, the simulations are evolved in a scale free form and the amplitude of the mass accretion rate is largely set by the density in the model.  We are primarily interested in the consistent qualitative behavior here, since all of the models are initialized with the same density normalization.  This means the integrated ``mass" of the disk is larger with increasing scaleheight, which, when coupled with the shorter evolutionary timescales and shorter accretion timescales, should give higher $\dot{M}$ with progressively larger scaleheights.

\subsection{Observational Signatures}
\label{sec-obs_sigs}

On their own, the distinct dynamo behaviors we find are interesting, but greater physical meaning can be found by connecting the unique manifestations to observables.  The rich photometric variability seen from accreting black holes encodes information about the accretion process, with the imprint of the dynamo being a likely component of this signal.  Since we neglect detailed radiative physics in order to save computational resources, we explore this with an emission proxy.  We use a scheme employed by other global accretion disk simulations (e.g. \citet{2001ApJ...548..348H}, \citet{2003MNRAS.341.1041A} \& \citet{2016ApJ...826...40H}) based on the internal disk stress.  Adopting Eqn 9 from \citet{1998ApJ...505..558H}, the local flux at the photosphere of the disk to dissipation is given by 
\begin{equation}
F=\frac{3}{2}\sqrt{\frac{GM}{r^{3}}}\Bigg(\frac{A}{B}\Bigg) \int_{0}^{h} B_{r}B_{\phi} dz,
\end{equation} where A and B are relativistic correction factors given by 
\begin{equation}
A=1-\frac{2GM}{r c^2}
\end{equation} and
\begin{equation}
B=1-\frac{3GM}{r c^2}.
\end{equation} For each data dump we integrate $F$ within $r=25\:r_g$ to calculate a bolometric ``luminosity," $L$.  The region beyond $r=25\:r_g$ is excluded because of residual transient behavior from the initialization of the disk which affects the very early part of our analysis phase.  This also contaminates the very early part of the \texttt{hr\_005} lightcurve, which we ignore in our calculation of its standard deviation.  We normalize by the mean luminosity to express the variability in a fractional form and because the models are evolved dimensionless, scale free form.  The lightcurves are shown in Figure \ref{fig-lightcurves}.

The most obvious difference in the synthetic light curves is the level of organization provided by the relative coherence of the dynamo.  Prior numerical studies have shown the stress in the disk is modulated by the dynamo \citep{2010ApJ...713...52D, 2012ApJ...744..144F, 2016ApJ...826...40H}, which provides a link to the disk heating as the energy injected by the dynamo is ultimately deposited as heat.  In our simulations, we see that the thinner disks with a more organized dynamo display slower undulations and a smaller fractional amplitude.  As measured by the standard deviation, \begin{equation}
\sigma = \sqrt{\frac{1}{N} \sum_{i=1}^N \Big(L_i - \overline{L}\Big)^2},
\end{equation} the typical fractional amplitude of the \texttt{hr\_005} model is $\sigma=0.14$ and for the \texttt{hr\_01} model it is $\sigma=0.20$. 
 
The thicker disks display rapid, incoherent variability with a much larger fractional amplitude.  In the \text{hr\_02} model, $\sigma=0.24$ and in the \texttt{h\_04} model, $\sigma=0.35$.  In these models, the fluctuations are much more ``flarey" and the light curves show rapid brightening episodes and subsequent troughs.  For instance, in the \texttt{hr\_02} model from $t=1.1\times10^4\:GM/c^3$ to $t=1.3\times10^4\:GM/c^3$ when the disk luminosity would appear to diminish by $36\%$, only to quickly rebrighten by a factor of three.  Later, from $t=1.7\times10^4GM/c^3$ to $t=1.9\times10^4\:GM/c^3$, the disk dips to half its baseline value, only to return quickly.  The \texttt{hr\_04} model shows less structure than the \texttt{hr\_02} model, with fast, stochastic fluctuations of roughly factors of $2.5$, i.e. fluctuations between $0.6$ and $1.5$ on several occasions.  At two points there are large excursions from the mean with flares that peak at over twice the mean value.

\section{Discussion}
\label{sec-discussion}

The results from the suite of accretion disk simulations we present here highlight how poorly the dynamo mechanism in accretion disks is understood and the additional work that is needed to fully leverage this phenomenon as an observational probe.  The crux of this work is that the large-scale dynamo can fail to organize into the low-frequency, quasiperiodic butterfly pattern typically seen in time traces of the vertical, azimuthally averaged toroidal magnetic field ($\langle B_\phi \rangle$).  Instead, in the thicker accretion disks the field is amplified in a stochastic way and on small-scales with no obvious structure or order, but still at low frequencies.  This challenges the ubiquity of a self-organized large-scale dynamo appearing as a feature of black hole disks.  Consequently, this could translate into observable signatures as the synthetic lightcurves we generate reflect the degree of order in the magnetic field.  

Unfortunately, we cannot pin down a detailed reason for why the dynamo fails to organize when the disk thickness is increased.  However, these results suggest a separation between turbulent fluctuating scales and the dominant scale of the flow is needed for the dynamo organization.  Considering the MHD turbulence spectrally has led to ``shell" models of the interactions in Fourier space where it has been shown that the magnetic and velocity fluctuations both decay locally to larger wavenumber.  However, the interactions between the two occur non-locally \citep{1978mfge.book.....M, 2011AnRFM..43..377M} as long wavelength velocity fluctuations are needed to stretch the magnetic field lines and long wavelength magnetic field fluctuations are needed to apply a net Lorentz force.  Understanding the details of how the spectral behaviors and interactions of the turbulence depend on disk thickness may help to tie the development of the large-scale dynamo into this framework.

\begin{figure*}
\includegraphics{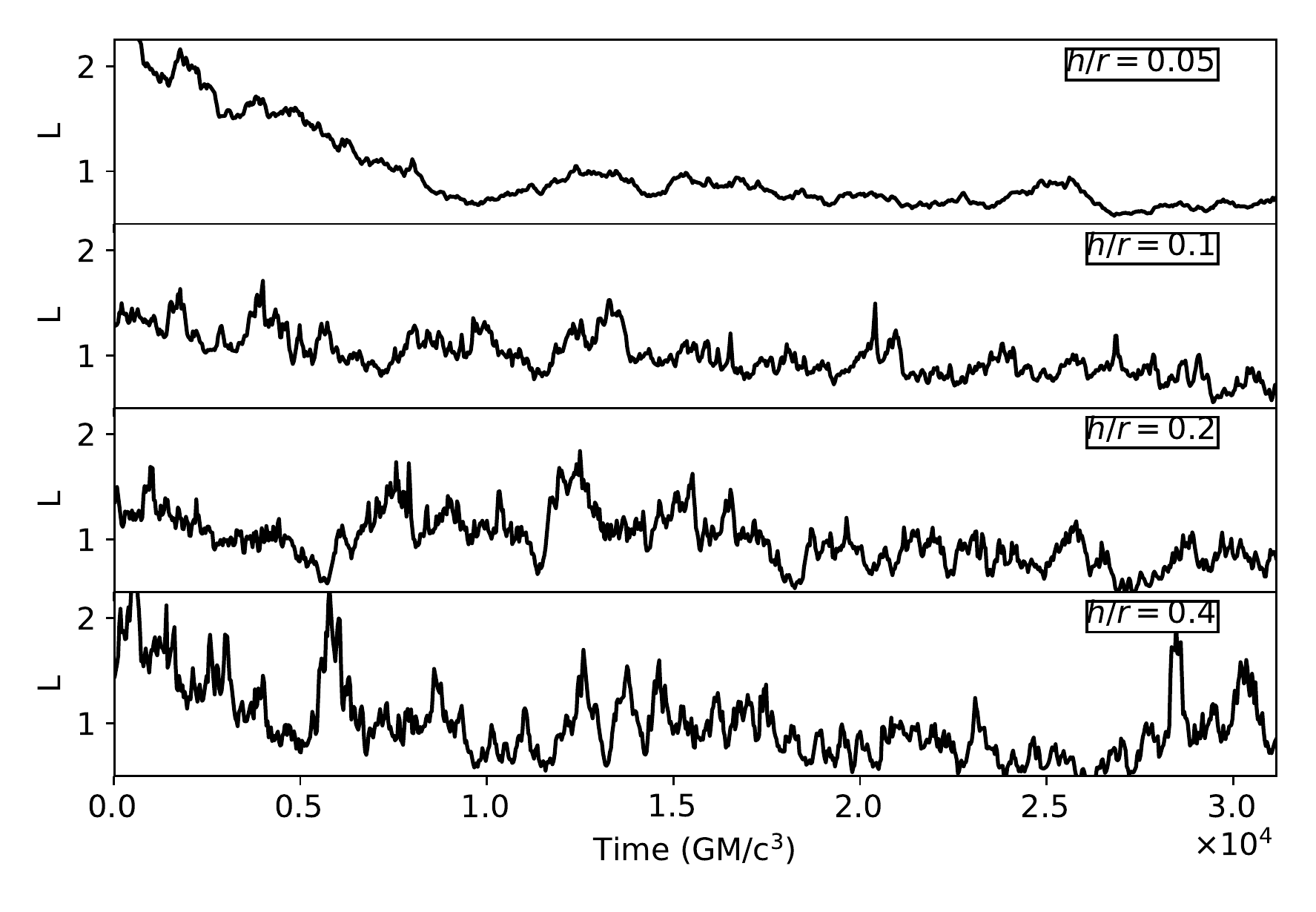}
\caption{Synthetic lightcurves calculated from the disk cooling for \texttt{hr\_005} (top),  \texttt{hr\_01} (second to top),  \texttt{hr\_02} (second to bottom), and  \texttt{hr\_04} (bottom).  The cooling has been normalized to the mean value of the lightcurve to express the amplitude in terms of a fractional amplitude.
\label{fig-lightcurves}}
\end{figure*}

Related to this, the different turbulent scales may change the \emph{effective} magnetic Prandtl number, Pr$_m$.  This could be a viable explanation since the effective diffusivity changes because of the larger turbulent eddies, which we find evidence for here in the increased scatter of the $\langle \mathcal{E'_{\phi}} \rangle$ and $\langle B_\phi \rangle$ correlations.  \citet{2009ApJ...697.1901G} explored the role of the turbulent Pr$_m$ in MHD accretion disks, and found it increased with increasing scaleheight.  The excitation of the MRI has been shown to depend on Pr$_{m}$ \citep{2007A&A...476.1113F, 2007A&A...476.1123F}, and simulations of forced turbulence have shown that the large-scale dynamo excitation also has a dependence on Pr$_{m}$, depending on how the turbulence is injected \citep{2014ApJ...791...12B}.  Since the effective viscosity, as measured by $\alpha_{ss}$, is roughly constant, this could show the role of the larger eddies increasing the turbulent magnetic diffusivity and connect to disappearance of they dynamo cycle at low Pr$_m$ \citep{2015A&A...575A..14R}.

Investigating the role of helicity in the disk may help in deciphering the differences between the thick and thin disks.  The evolution and influence of the flow helicities is expected to be intimately linked to global field behavior, and could be important in the excitation and organization of a large scale dynamo \citep[e.g.][]{2006PhRvE..74f6310F, 2011MNRAS.413..901K, 2012PhyS...86e8202B, 2015SSRv..188...59B}.  However, it is unclear the role that it plays in the global behavior of an accretion disk.  The clear case of a relation between the large-scale magnetic helicity with the azimuthally averaged toroidal field in the \texttt{hr\_005} simulation offers a glimpse into the the behavior.  Pursuing this connection more may help to explain the curious changes in the parity between the two disk hemispheres, as well as abnormal features like the intermittency and ``failed reversals" in the field oscillations because they may be related to the ability of the disk to assemble the large-scale field into an ordered pattern and its ability to shed the helicity generated by the turbulence.

Fundamentally, all of these scenarios are related and could all be contributing to the disorganization of the dynamo on some level.  Moving forward, there are several routes that could clarify why the large scale dynamo failed to organize.  Using test field methods, first applied to the geodynamo \citep{2005AN....326..245S, 2007GApFD.101...81S}, to measure the mean-field dynamo coefficients in these global runs is possibility the most important next step as it has been effectively used to characterize the dynamo in local shearing box simulations \citep{2010MNRAS.405...41G, 2015ApJ...810...59G}.  Additionally, expanding the simulations to encompass the full $2\pi$ azimuthal domain could be significant for fully capturing of the large scale structure.  To run the simulations for our planned duration, we were restricted to a truncated azimuthal domain out of necessity.  Our simulations were already very computational expensive, with \texttt{hr\_01} requiring 2 million CPU hours and \texttt{hr\_005} requiring over 7 million CPU hours, so extending the domain with proper resolution like we had in these models was impractical.  A number of dedicated resolution studies have shown that the properties of the MRI turbulence can depend on the domain and it is possible we are simply seeing an analogous sensitivity in the dynamo.  In Appendix \ref{sec-appendix} we present two test simulations of \texttt{hr\_02} and \texttt{hr\_04} that have $\phi$-domains extended to $\Delta\phi=\pi$.  These simulations show identical behavior to their compliments used in this analysis, which offers encouraging evidence the grid design is not the root cause of the unstructured pattern.  This bolsters these results, but subtle numerical effects remain a lingering concern.  Guided by this first attempt to explore the scaleheight dependence of the dynamo, we can tailor our simulations to target the two interesting regimes.  Furthermore, we were intentionally very conservative with the length of the initialization phase in these simulations to prevent transients from distorting our results, and it could likely be shorted to redirect computational resources in the future.

As a note, in \citet{2018ApJ...854....6H} we studied an MHD model of a ``truncated" accretion disk where there is believed to be a transition region between a hot, radiatively inner accretion flow and a cooler, radiatively efficient disk.  In the outer thin disk ($h/r=0.1$), we find the large scale dynamo readily develops and is sustained.  However, in inner region where the disk thickens ($h/r>0.2$), we find it fails to develop, similar to this study.  In the truncated disk the flow had the added component of an outflow originating from the truncation zone, so it is not an isolated system like those presented here, but the failed organization of the dynamo in the thicker disk is reminiscent of that detailed in this paper.  The truncated disk scenario essentially provides a composite accretion flow and demonstrates that it truly is a scale height dependence since the two distinct flow behaviors develop in close proximity and can even interact.

The observational consequences of changing disk height are of great interest.  The lightcurves we present in Figure \ref{fig-lightcurves} show that the temporal behavior of the photometric variability from emission should be distinct, which meshes with empirical results.  Accreting stellar mass black holes in black hole binaries (BHBs) and supermassive black holes in active galactic nuclei (AGNs) display a bifurcation in spectral states that is typically attributed to the radiative efficiency of the system.  In the low-hard state of BHBs and in low-luminosity AGNs (LLAGNs) the accretion flow is presumed to be radiatively inefficient and hot.  The accretion flow should, therefore, take on a thicker disk geometry since it cannot radiate away its thermal energy.  High-soft state BHBs and the typical Seyfert-like AGNs and quasars, on the other hand, should be able to radiate efficiently, so the accretion is expected to occur through a thin disk.

Making a direct association between the lightcurves from our simulations to astrophysical black hole systems is not straightforward since the disk emission is produced by different processes and in different wavebands for the thin and thick disk cases, but it nevertheless seems to hold.  Indeed, this scheme matches the trends from BHBs as they change states during outburst \citep{2006ARA&A..44...49R}.  Variability in AGNs is poorly understood, but the different timing properties of LLAGNs compared to Seyfert-like AGNs may be a clue about their ability to host a large-scale dynamo in the disk.  LLAGNs typically have rapid variability e.g. NGC 4258 \citep{2005ApJ...625L..39M} and NGC 3226 \citep{2009ApJ...691..431B}, but they are less well studied their higher Eddington ratio counterparts.  Extensive monitoring campaigns have been completed across the electromagnetic spectrum for Seyfert-like AGNs which show variability on a thermal time, $t_{therm}\approx 1/\alpha\Omega$,  \citep{2009ApJ...698..895K, 2010ApJ...721.1014M, 2017MNRAS.470.3027K}, roughly the same timescale that the dynamo oscillations might present themselves.

The results we present here may have additional observational impacts as a number of other variable processes could stem from the presence of a well-ordered large scale dynamo.  Features like dynamo driven low-frequency QPOs could be impacted by the departure from the assumed cyclical behavior, so it is prudent to revisit their utility as probes of the central black hole mass and spin \citep[e.g.][]{2008Natur.455..369G, 2010ApJ...710...16Z, 2014Natur.513...74P, 2016ApJ...819L..19P} to verify the assumptions that serve as the foundation of the mass scaling remain valid since they will not appear if they dynamo is unordered.  

In \citet{2016ApJ...826...40H} we presented a detailed analysis of a simulated thin disk where propagating fluctuations in mass accretion rate naturally developed from the turbulence.  At its core, the telling nonlinear signatures that are commonly observed, i.e. log-normal flux distribution, linear relations between the RMS and flux level of the variability, and interband coherence where the harder emission lags the softer emission, arise from the multiplicative combination of stochastic fluctuations in the mass accretion rate.  As we emphasized the name ``propagating fluctuations" is a misnomer since the phenomenon is fundamentally just the preservation of the fluctuation pattern as angular momentum is diffusively redistributed according to the canonical disk equation \citep{1981ARA&A..19..137P}, \begin{equation} \label{eqn-can_disk}
\frac{\partial \Sigma}{\partial t} = \frac{3}{R}\frac{\partial}{\partial R}\Big[R^{\frac{1}{2}} \frac{\partial}{\partial R}(\nu \Sigma R^{\frac{1}{2}})\Big].
\end{equation}  A key component in the growth of propagating fluctuations in mass accretion rate is that the effective viscosity must be modulated at low enough frequencies that turbulent fluctuations do not wipe out the growth of the structure in the accretion flow \citep{2014ApJ...791..126C}.  The results presented here show that this condition is met, even when the dynamo is not ordered into its standard cycle because low-frequency power is still seen in the 2D PSDs.  The log-normal $\dot{M}$ distributions we find in these simulations further confirms that the growth of propagating fluctuations is hardy enough to occur in an accretion disk regardless of the dynamo organization, the dynamo just simply needs operate.

There has been a recent push to understand the dynamo beyond the standard thin accretion disk, and from these efforts a narrative is developing that it is possible to impede or alter the growth of the large-scale dynamo.  Here, we present one more example of modification of the large-scale dynamo.  Further investigation into the peculiarities of the dynamo are needed and warranted to get a better handle on when it is a reliable source of variability and when it cannot contribute to the observational signatures.  This discussion is meant to highlight several ways in which commonly observed features could be impacted by the breakdown of the dynamo organization, but it is by no means complete.  There are additional ramifications beyond these that could be significant for interpreting the accretion flow dynamics around a black hole.

\section{Conclusion}
\label{sec-conclusion}

With this study, we sought to clarify how the dynamo fits into the larger puzzle of black hole accretion.  Using a suite of four global, MHD accretion disk simulations with scale height ratios $h/r=\{0.05, 0.1, 0.2,0.4\}$, we expose a scaleheight dependence in the ability of an accretion disk to organize the large-scale dynamo.  In summary, our top-line results from these simulations are:
\begin{enumerate}

\item Low-frequency, ordered oscillations in the azimuthal magnetic field from the large-scale dynamo are present in the \texttt{hr\_005} ($h/r=0.05$) and \texttt{hr\_01} ($h/r=0.1$) models, but are increasingly absent in the \texttt{hr\_02} ($h/r=0.2$) and \texttt{hr\_04} ($h/r=0.4$) models.
\item When the organized large-scale dynamo is present in the thinner disks, there is a coherent band of power in the PSD of azimuthally average toroidal magnetic field, $\langle B_\phi \rangle$, at approximately $10\times$ the local orbital period.  In the thicker accretion disks where the large-scale dynamo is unorganized, the PSD is featureless with power on all timescales, down the the lowest frequencies we can probe with the duration of our simulations.
\item Calculation of $\alpha_{d}$ through correlations between $\langle B_\phi \rangle$ and the turbulent electromotive force, $\langle \mathcal{E'}_\phi \rangle$, in the coronal regions of the simulations yield similar values across our models.  In the upper hemisphere of the simulations $\alpha_{d}$ is negative, while in the lower hemisphere it is positive.
\item In synthetic light curves produced through a proxy, the presence of a large-scale dynamo is related to the level of order and amplitude of the fluctuations.  The light curves of the thicker disks display large amplitude, stochastic fluctuations, which reflects the lack of organization in the accretion flow.  The thinner disk simulations, in comparison, show a dearth of large amplitude variation and smaller-scale, slower undulations instead.

\end{enumerate}

As we continue to assemble the accretion puzzle and interpret variability from accreting black holes, exploring the details of the underlying physics is an important pursuit.  The spectrotemporal behavior provides a crucial window into these systems, but a first principles understanding of their origin remains elusive.  This hinders the ability to leverage the observational signatures to their full potential and offers an opportunity for future advancement.  The odd dynamo behavior we detail here has the immediate implication that the large-scale dynamo shows markedly different appearance in thicker accretion disks compared to its well characterized behavior in thin disks, which could have broader importance in the global accretion disk evolution.
\vspace{-0.5cm}
\acknowledgements

We thank Omer Blaes, Christian Knigge, and Oliver Gressel for the useful discussions.  The authors also thank the anonymous referee for useful comments that helped strengthen and clarify the paper.  JDH thanks support from NASA under the NASA Earth and Space Science Fellowship program (grant NNX16AP88H).  The authors acknowledge the University of Maryland supercomputing resources (http://hpcc.umd.edu), including the Deepthought2, Deepthought, and MARCC/Bluecrab clusters, made available for conducting the research reported in this paper.

 \appendix
 \section{Extended Domain Tests}
 \label{sec-appendix}
 
To confirm the results from our analysis, we ran two additional simulations of the thickest accretion disks, \texttt{hr\_02\_pi} and \texttt{hr\_04\_pi}.  These simulations were constructed identically to their $\Delta \phi=\pi/3$ counterparts, but with a $\phi$-domain extended to $\Delta\phi=\pi$.  The resolution element is preserved by tripling the number of zones in the azimuthal direction so that \texttt{hr\_02\_pi} has $N_{R} \times N_{\theta} \times N_{\phi} = 308 \times 248 \times 192$ zones and \texttt{hr\_04\_pi} has $N_{R} \times N_{\theta} \times N_{\phi} = 154 \times 248 \times 96$ zones.  The \texttt{hr\_02\_pi} model was integrated for $680$ ISCO orbits, or $t=4.19\times10^{4}\:GM/c^3$. The \texttt{hr\_02\_pi} model was also integrated for $680$ ISCO orbits, or $t=4.19\times10^{4}\:GM/c^3$.  As before, only the final $\Delta t = 3.15\times10^4\:GM/c^3$ ($512$ ISCO orbits) is used in the analysis.
 
  \begin{figure*}[!btp]
\includegraphics{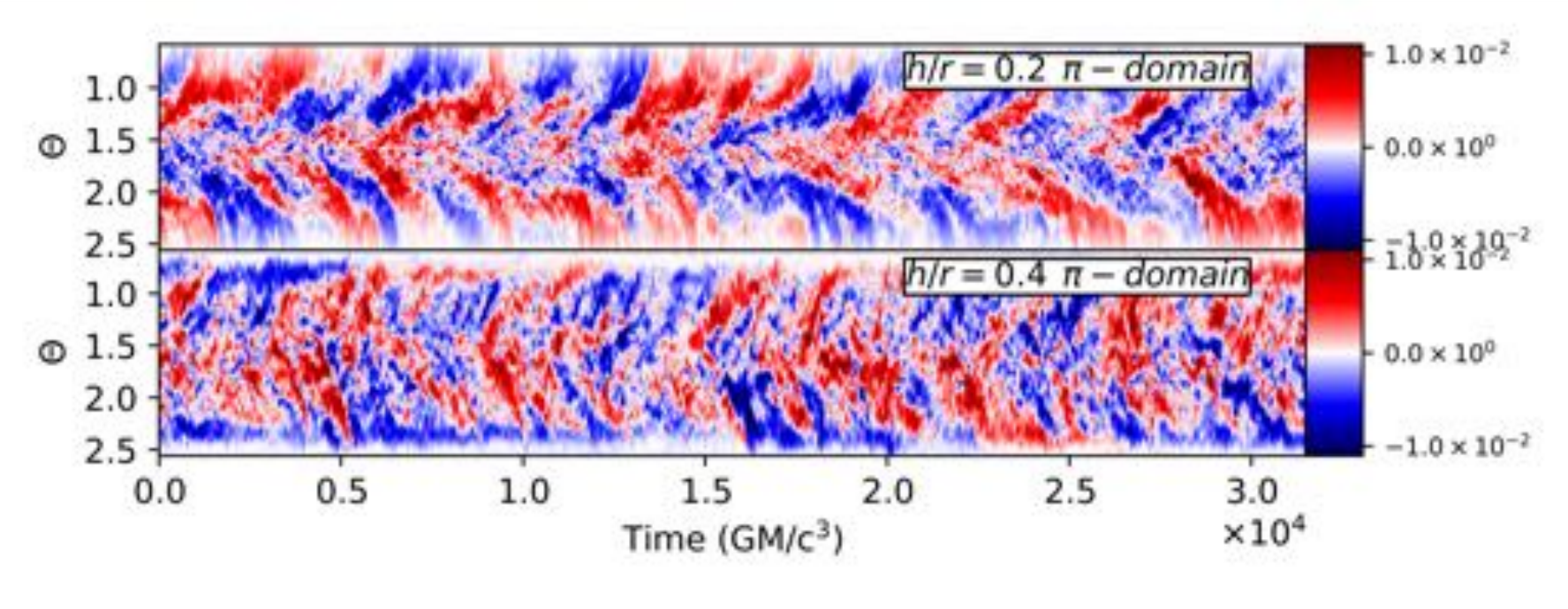}
\caption{Spacetime diagrams of the azimuthally averaged $B_\phi$ at $r=15\:r_g$ for the extended $\pi$-domain \texttt{hr\_02} (top) and \texttt{hr\_04} (bottom) tests.  Positive values (red) indicate orientation of the field in the positive $\phi$-direction while negative (blue) indicates an opposite orientation.  Color intensity corresponds to the averaged magnitude.
\label{fig-butterfly_diagrams_appendix}}
\end{figure*}

\begin{figure*}[!btp]
\includegraphics{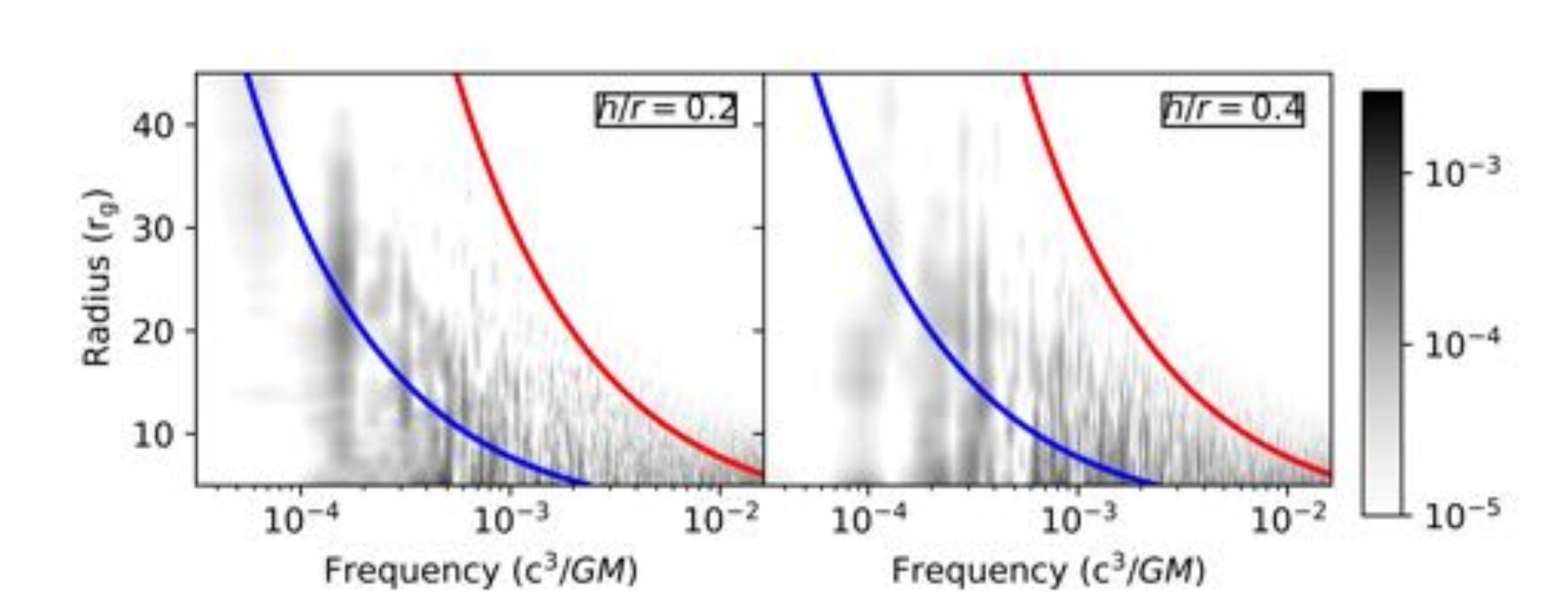}
\caption{PSDs of $B_{\phi}$ at $1.5h$ above the disk midplane for the extended $\pi$-domain \texttt{hr\_02} (left), and \texttt{hr\_04} (right) tests.  Darker colors (black) represents greater power in that frequency bin for a given radius.  The orbital frequency is shown with the red line and  ten times the orbital frequency is shown with the blue line.
\label{fig-2D_FFT_appendix}}
\end{figure*}
 
Extending the $\phi$-domain places these models within a similar regime to \texttt{hr\_005} and \texttt{hr\_01} in terms of the number of disk scale heights per azimuthal expanse.  Since the large-scale dynamo depends on the self-organization of the MRI-driven turbulence, a smaller domain might not fit the global flow structure that leads to this behavior.  However, these models would allow similar butterfly patterns to develop in these thicker disk simulations if our previous results were solely due to a failure to capture the global modes.
 
Figure \ref{fig-butterfly_diagrams_appendix} shows spacetime diagrams of the azimuthally averaged $B_\phi$ of \texttt{hr\_02\_pi} and \texttt{hr\_04\_pi} at $r=15\:r_g$.  As seen in \texttt{hr\_02}, the dynamo attempts to establish a butterfly pattern, but it is much more irregular and disorganized than that seen in the \texttt{hr\_005} and \texttt{hr\_01} models.  Sometimes an organized field dominates for a long portion of the simulation, i.e. $t=1.3-1.5\times10^4\:GM/c^3$ in the upper hemisphere of the simulation, while at others it is only momentary present.  The \texttt{hr\_04\_pi} simulations shows no organization, like \texttt{hr\_04}.  Fluctuations in the field are rapid, disorganized, and chaotic which resembles nothing more than turbulent fluctuations.  Figure \ref{fig-2D_FFT_appendix} shows the PSDs of  \texttt{hr\_02\_pi} and \texttt{hr\_04\_pi}.  Like Figure \ref{fig-2D_FFT}, the thick disks show broad bands of power with no evidence of the typical band of power found at one tenth the orbital frequency.   The similarity between the PSDs offers secondary evidence the behavior of the magnetic field evolution does not depend on the domain size.
 
For this analysis, these $\Delta\phi=\pi$ simulations offer compelling evidence that the breakdown in the dynamo pattern with increasing disk scaleheight is a real effect and not a nuance that arises from the grid scheme.  In Section \ref{sec-discussion} we offer several refinements for future studies that will help alleviate any remaining concern of numerical artifacts.  Incorporating these improvements with additional features, like a test-field method to probe the field evolution, will provide valuable insights into the dynamo process.
\newpage
\bibliography{dynamo_bib}

\begin{thebibliography}{}
\expandafter\ifx\csname natexlab\endcsname\relax\def\natexlab#1{#1}\fi

\bibitem[{{Armitage} \& {Reynolds}(2003)}]{2003MNRAS.341.1041A}
{Armitage}, P.~J., \& {Reynolds}, C.~S. 2003, \mnras, 341, 1041

\bibitem[{{Bai} \& {Stone}(2013)}]{2013ApJ...767...30B}
{Bai}, X.-N., \& {Stone}, J.~M. 2013, \apj, 767, 30

\bibitem[{{Balbus} {et~al.}(1994){Balbus}, {Gammie}, \&
  {Hawley}}]{1994MNRAS.271..197B}
{Balbus}, S.~A., {Gammie}, C.~F., \& {Hawley}, J.~F. 1994, \mnras, 271,
  doi:10.1093/mnras/271.1.197

\bibitem[{{Balbus} \& {Hawley}(1991)}]{1991ApJ...376..214B}
{Balbus}, S.~A., \& {Hawley}, J.~F. 1991, \apj, 376, 214

\bibitem[{{Beckwith} {et~al.}(2011){Beckwith}, {Armitage}, \&
  {Simon}}]{2011MNRAS.416..361B}
{Beckwith}, K., {Armitage}, P.~J., \& {Simon}, J.~B. 2011, \mnras, 416, 361

\bibitem[{{Begelman} {et~al.}(2015){Begelman}, {Armitage}, \&
  {Reynolds}}]{2015ApJ...809..118B}
{Begelman}, M.~C., {Armitage}, P.~J., \& {Reynolds}, C.~S. 2015, \apj, 809, 118

\bibitem[{{Binder} {et~al.}(2009){Binder}, {Markowitz}, \&
  {Rothschild}}]{2009ApJ...691..431B}
{Binder}, B., {Markowitz}, A., \& {Rothschild}, R.~E. 2009, \apj, 691, 431

\bibitem[{{Blackman}(2012)}]{2012PhyS...86e8202B}
{Blackman}, E.~G. 2012, \physscr, 86, 058202

\bibitem[{{Blackman}(2015)}]{2015SSRv..188...59B}
---. 2015, \ssr, 188, 59

\bibitem[{{Blackman} \& {Brandenburg}(2002)}]{2002ApJ...579..359B}
{Blackman}, E.~G., \& {Brandenburg}, A. 2002, \apj, 579, 359

\bibitem[{{Bodo} {et~al.}(2008){Bodo}, {Mignone}, {Cattaneo}, {Rossi}, \&
  {Ferrari}}]{2008A&A...487....1B}
{Bodo}, G., {Mignone}, A., {Cattaneo}, F., {Rossi}, P., \& {Ferrari}, A. 2008,
  \aap, 487, 1

\bibitem[{{Brandenburg}(2014)}]{2014ApJ...791...12B}
{Brandenburg}, A. 2014, \apj, 791, 12

\bibitem[{{Brandenburg} \& {Donner}(1997)}]{1997MNRAS.288L..29B}
{Brandenburg}, A., \& {Donner}, K.~J. 1997, \mnras, 288, L29

\bibitem[{{Brandenburg} {et~al.}(1995){Brandenburg}, {Nordlund}, {Stein}, \&
  {Torkelsson}}]{1995ApJ...446..741B}
{Brandenburg}, A., {Nordlund}, A., {Stein}, R.~F., \& {Torkelsson}, U. 1995,
  \apj, 446, 741

\bibitem[{{Brandenburg} \& {Subramanian}(2005)}]{2005PhR...417....1B}
{Brandenburg}, A., \& {Subramanian}, K. 2005, \physrep, 417, 1

\bibitem[{{Chandrasekhar}(1960)}]{1960PNAS...46..253C}
{Chandrasekhar}, S. 1960, Proceedings of the National Academy of Science, 46,
  253

\bibitem[{{Coleman} {et~al.}(2017){Coleman}, {Yerger}, {Blaes}, {Salvesen}, \&
  {Hirose}}]{2017MNRAS.467.2625C}
{Coleman}, M.~S.~B., {Yerger}, E., {Blaes}, O., {Salvesen}, G., \& {Hirose}, S.
  2017, \mnras, 467, 2625

\bibitem[{{Cowperthwaite} \& {Reynolds}(2014)}]{2014ApJ...791..126C}
{Cowperthwaite}, P.~S., \& {Reynolds}, C.~S. 2014, \apj, 791, 126

\bibitem[{{Davis} {et~al.}(2010){Davis}, {Stone}, \&
  {Pessah}}]{2010ApJ...713...52D}
{Davis}, S.~W., {Stone}, J.~M., \& {Pessah}, M.~E. 2010, \apj, 713, 52

\bibitem[{{Flock} {et~al.}(2010){Flock}, {Dzyurkevich}, {Klahr}, \&
  {Mignone}}]{2010A&A...516A..26F}
{Flock}, M., {Dzyurkevich}, N., {Klahr}, H., \& {Mignone}, A. 2010, \aap, 516,
  A26

\bibitem[{{Flock} {et~al.}(2012){Flock}, {Dzyurkevich}, {Klahr}, {Turner}, \&
  {Henning}}]{2012ApJ...744..144F}
{Flock}, M., {Dzyurkevich}, N., {Klahr}, H., {Turner}, N., \& {Henning}, T.
  2012, \apj, 744, 144

\bibitem[{{Frick} {et~al.}(2006){Frick}, {Stepanov}, \&
  {Sokoloff}}]{2006PhRvE..74f6310F}
{Frick}, P., {Stepanov}, R., \& {Sokoloff}, D. 2006, \pre, 74, 066310

\bibitem[{{Fromang} \& {Papaloizou}(2007)}]{2007A&A...476.1113F}
{Fromang}, S., \& {Papaloizou}, J. 2007, \aap, 476, 1113

\bibitem[{{Fromang} {et~al.}(2007){Fromang}, {Papaloizou}, {Lesur}, \&
  {Heinemann}}]{2007A&A...476.1123F}
{Fromang}, S., {Papaloizou}, J., {Lesur}, G., \& {Heinemann}, T. 2007, \aap,
  476, 1123

\bibitem[{{Gierli{\'n}ski} {et~al.}(2008){Gierli{\'n}ski}, {Middleton}, {Ward},
  \& {Done}}]{2008Natur.455..369G}
{Gierli{\'n}ski}, M., {Middleton}, M., {Ward}, M., \& {Done}, C. 2008, \nat,
  455, 369

\bibitem[{{Gogichaishvili} {et~al.}(2017){Gogichaishvili}, {Mamatsashvili},
  {Horton}, {Chagelishvili}, \& {Bodo}}]{2017arXiv170707044G}
{Gogichaishvili}, D., {Mamatsashvili}, G., {Horton}, W., {Chagelishvili}, G.,
  \& {Bodo}, G. 2017, ArXiv e-prints, arXiv:1707.07044

\bibitem[{{Gressel}(2010)}]{2010MNRAS.405...41G}
{Gressel}, O. 2010, \mnras, 405, 41

\bibitem[{{Gressel} \& {Pessah}(2015)}]{2015ApJ...810...59G}
{Gressel}, O., \& {Pessah}, M.~E. 2015, \apj, 810, 59

\bibitem[{{Guan} \& {Gammie}(2009)}]{2009ApJ...697.1901G}
{Guan}, X., \& {Gammie}, C.~F. 2009, \apj, 697, 1901

\bibitem[{{Guan} \& {Gammie}(2011)}]{2011ApJ...728..130G}
---. 2011, \apj, 728, 130

\bibitem[{{Guan} {et~al.}(2009){Guan}, {Gammie}, {Simon}, \&
  {Johnson}}]{2009ApJ...694.1010G}
{Guan}, X., {Gammie}, C.~F., {Simon}, J.~B., \& {Johnson}, B.~M. 2009, \apj,
  694, 1010

\bibitem[{{Hawley} {et~al.}(1995){Hawley}, {Gammie}, \&
  {Balbus}}]{1995ApJ...440..742H}
{Hawley}, J.~F., {Gammie}, C.~F., \& {Balbus}, S.~A. 1995, \apj, 440, 742

\bibitem[{{Hawley} {et~al.}(2011){Hawley}, {Guan}, \&
  {Krolik}}]{2011ApJ...738...84H}
{Hawley}, J.~F., {Guan}, X., \& {Krolik}, J.~H. 2011, \apj, 738, 84

\bibitem[{{Hawley} \& {Krolik}(2001)}]{2001ApJ...548..348H}
{Hawley}, J.~F., \& {Krolik}, J.~H. 2001, \apj, 548, 348

\bibitem[{{Hawley} {et~al.}(2013){Hawley}, {Richers}, {Guan}, \&
  {Krolik}}]{2013ApJ...772..102H}
{Hawley}, J.~F., {Richers}, S.~A., {Guan}, X., \& {Krolik}, J.~H. 2013, \apj,
  772, 102

\bibitem[{{Hawley} \& {Stone}(1995)}]{1995CoPhC..89..127H}
{Hawley}, J.~F., \& {Stone}, J.~M. 1995, Computer Physics Communications, 89,
  127

\bibitem[{{Hogg} \& {Reynolds}(2016)}]{2016ApJ...826...40H}
{Hogg}, J.~D., \& {Reynolds}, C.~S. 2016, \apj, 826, 40

\bibitem[{{Hogg} \& {Reynolds}(2018)}]{2018ApJ...854....6H}
---. 2018, \apj, 854, 6

\bibitem[{{Hubeny} \& {Hubeny}(1998)}]{1998ApJ...505..558H}
{Hubeny}, I., \& {Hubeny}, V. 1998, \apj, 505, 558

\bibitem[{{Jiang} {et~al.}(2014){Jiang}, {Stone}, \&
  {Davis}}]{2014ApJ...796..106J}
{Jiang}, Y.-F., {Stone}, J.~M., \& {Davis}, S.~W. 2014, \apj, 796, 106

\bibitem[{{K{\"a}pyl{\"a}} \& {Korpi}(2011)}]{2011MNRAS.413..901K}
{K{\"a}pyl{\"a}}, P.~J., \& {Korpi}, M.~J. 2011, \mnras, 413, 901

\bibitem[{{Kasliwal} {et~al.}(2017){Kasliwal}, {Vogeley}, \&
  {Richards}}]{2017MNRAS.470.3027K}
{Kasliwal}, V.~P., {Vogeley}, M.~S., \& {Richards}, G.~T. 2017, \mnras, 470,
  3027

\bibitem[{{Kelly} {et~al.}(2009){Kelly}, {Bechtold}, \&
  {Siemiginowska}}]{2009ApJ...698..895K}
{Kelly}, B.~C., {Bechtold}, J., \& {Siemiginowska}, A. 2009, \apj, 698, 895

\bibitem[{{King} {et~al.}(2004){King}, {Pringle}, {West}, \&
  {Livio}}]{2004MNRAS.348..111K}
{King}, A.~R., {Pringle}, J.~E., {West}, R.~G., \& {Livio}, M. 2004, \mnras,
  348, 111

\bibitem[{{Krause} \& {Raedler}(1980)}]{1980mfmd.book.....K}
{Krause}, F., \& {Raedler}, K.~H. 1980, {Mean-field magnetohydrodynamics and
  dynamo theory}

\bibitem[{{Krolik} \& {Hawley}(2002)}]{2002ApJ...573..754K}
{Krolik}, J.~H., \& {Hawley}, J.~F. 2002, \apj, 573, 754

\bibitem[{{Latter} {et~al.}(2009){Latter}, {Lesaffre}, \&
  {Balbus}}]{2009MNRAS.394..715L}
{Latter}, H.~N., {Lesaffre}, P., \& {Balbus}, S.~A. 2009, \mnras, 394, 715

\bibitem[{{Lesur} \& {Ogilvie}(2008)}]{2008A&A...488..451L}
{Lesur}, G., \& {Ogilvie}, G.~I. 2008, \aap, 488, 451

\bibitem[{{MacLeod} {et~al.}(2010){MacLeod}, {Ivezi{\'c}}, {Kochanek},
  {Koz{\l}owski}, {Kelly}, {Bullock}, {Kimball}, {Sesar}, {Westman}, {Brooks},
  {Gibson}, {Becker}, \& {de Vries}}]{2010ApJ...721.1014M}
{MacLeod}, C.~L., {Ivezi{\'c}}, {\v Z}., {Kochanek}, C.~S., {et~al.} 2010,
  \apj, 721, 1014

\bibitem[{{Markowitz} \& {Uttley}(2005)}]{2005ApJ...625L..39M}
{Markowitz}, A., \& {Uttley}, P. 2005, \apjl, 625, L39

\bibitem[{{Mayer} \& {Pringle}(2006)}]{2006MNRAS.368..379M}
{Mayer}, M., \& {Pringle}, J.~E. 2006, \mnras, 368, 379

\bibitem[{{Mignone} {et~al.}(2007){Mignone}, {Bodo}, {Massaglia}, {Matsakos},
  {Tesileanu}, {Zanni}, \& {Ferrari}}]{2007ApJS..170..228M}
{Mignone}, A., {Bodo}, G., {Massaglia}, S., {et~al.} 2007, \apjs, 170, 228

\bibitem[{{Mininni}(2011)}]{2011AnRFM..43..377M}
{Mininni}, P.~D. 2011, Annual Review of Fluid Mechanics, 43, 377

\bibitem[{{Moffatt}(1978)}]{1978mfge.book.....M}
{Moffatt}, H.~K. 1978, {Magnetic field generation in electrically conducting
  fluids}

\bibitem[{{Murphy} \& {Pessah}(2015)}]{2015ApJ...802..139M}
{Murphy}, G.~C., \& {Pessah}, M.~E. 2015, \apj, 802, 139

\bibitem[{{Noble} {et~al.}(2009){Noble}, {Krolik}, \&
  {Hawley}}]{2009ApJ...692..411N}
{Noble}, S.~C., {Krolik}, J.~H., \& {Hawley}, J.~F. 2009, \apj, 692, 411

\bibitem[{{Obergaulinger} {et~al.}(2009){Obergaulinger}, {Cerd{\'a}-Dur{\'a}n},
  {M{\"u}ller}, \& {Aloy}}]{2009A&A...498..241O}
{Obergaulinger}, M., {Cerd{\'a}-Dur{\'a}n}, P., {M{\"u}ller}, E., \& {Aloy},
  M.~A. 2009, \aap, 498, 241

\bibitem[{{Oishi} \& {Mac Low}(2011)}]{2011ApJ...740...18O}
{Oishi}, J.~S., \& {Mac Low}, M.-M. 2011, \apj, 740, 18

\bibitem[{{O'Neill} {et~al.}(2011){O'Neill}, {Reynolds}, {Miller}, \&
  {Sorathia}}]{2011ApJ...736..107O}
{O'Neill}, S.~M., {Reynolds}, C.~S., {Miller}, M.~C., \& {Sorathia}, K.~A.
  2011, \apj, 736, 107

\bibitem[{{Pan} {et~al.}(2016){Pan}, {Yuan}, {Yao}, {Zhou}, {Liu}, {Zhou}, \&
  {Zhang}}]{2016ApJ...819L..19P}
{Pan}, H.-W., {Yuan}, W., {Yao}, S., {et~al.} 2016, \apjl, 819, L19

\bibitem[{{Parkin} \& {Bicknell}(2013)}]{2013ApJ...763...99P}
{Parkin}, E.~R., \& {Bicknell}, G.~V. 2013, \apj, 763, 99

\bibitem[{{Pasham} {et~al.}(2014){Pasham}, {Strohmayer}, \&
  {Mushotzky}}]{2014Natur.513...74P}
{Pasham}, D.~R., {Strohmayer}, T.~E., \& {Mushotzky}, R.~F. 2014, \nat, 513, 74

\bibitem[{{Pessah}(2010)}]{2010ApJ...716.1012P}
{Pessah}, M.~E. 2010, \apj, 716, 1012

\bibitem[{{Pessah} {et~al.}(2007){Pessah}, {Chan}, \&
  {Psaltis}}]{2007ApJ...668L..51P}
{Pessah}, M.~E., {Chan}, C.-k., \& {Psaltis}, D. 2007, \apjl, 668, L51

\bibitem[{{Pessah} \& {Goodman}(2009)}]{2009ApJ...698L..72P}
{Pessah}, M.~E., \& {Goodman}, J. 2009, \apjl, 698, L72

\bibitem[{{Pringle}(1981)}]{1981ARA&A..19..137P}
{Pringle}, J.~E. 1981, \araa, 19, 137

\bibitem[{{Remillard} \& {McClintock}(2006)}]{2006ARA&A..44...49R}
{Remillard}, R.~A., \& {McClintock}, J.~E. 2006, \araa, 44, 49

\bibitem[{{Riols} {et~al.}(2015){Riols}, {Rincon}, {Cossu}, {Lesur}, {Ogilvie},
  \& {Longaretti}}]{2015A&A...575A..14R}
{Riols}, A., {Rincon}, F., {Cossu}, C., {et~al.} 2015, \aap, 575, A14

\bibitem[{{Salvesen} {et~al.}(2016){Salvesen}, {Simon}, {Armitage}, \&
  {Begelman}}]{2016MNRAS.457..857S}
{Salvesen}, G., {Simon}, J.~B., {Armitage}, P.~J., \& {Begelman}, M.~C. 2016,
  \mnras, 457, 857

\bibitem[{{Sano} {et~al.}(2004){Sano}, {Inutsuka}, {Turner}, \&
  {Stone}}]{2004ApJ...605..321S}
{Sano}, T., {Inutsuka}, S.-i., {Turner}, N.~J., \& {Stone}, J.~M. 2004, \apj,
  605, 321

\bibitem[{{Schrinner} {et~al.}(2005){Schrinner}, {R{\"a}dler}, {Schmitt},
  {Rheinhardt}, \& {Christensen}}]{2005AN....326..245S}
{Schrinner}, M., {R{\"a}dler}, K.-H., {Schmitt}, D., {Rheinhardt}, M., \&
  {Christensen}, U. 2005, Astronomische Nachrichten, 326, 245

\bibitem[{{Schrinner} {et~al.}(2007){Schrinner}, {R{\"a}dler}, {Schmitt},
  {Rheinhardt}, \& {Christensen}}]{2007GApFD.101...81S}
{Schrinner}, M., {R{\"a}dler}, K.-H., {Schmitt}, D., {Rheinhardt}, M., \&
  {Christensen}, U.~R. 2007, Geophysical and Astrophysical Fluid Dynamics, 101,
  81

\bibitem[{{Shakura} \& {Sunyaev}(1973)}]{1973A&A....24..337S}
{Shakura}, N.~I., \& {Sunyaev}, R.~A. 1973, \aap, 24, 337

\bibitem[{{Simon} {et~al.}(2012){Simon}, {Beckwith}, \&
  {Armitage}}]{2012MNRAS.422.2685S}
{Simon}, J.~B., {Beckwith}, K., \& {Armitage}, P.~J. 2012, \mnras, 422, 2685

\bibitem[{{Simon} {et~al.}(2011){Simon}, {Hawley}, \&
  {Beckwith}}]{2011ApJ...730...94S}
{Simon}, J.~B., {Hawley}, J.~F., \& {Beckwith}, K. 2011, \apj, 730, 94

\bibitem[{{Sorathia} {et~al.}(2012){Sorathia}, {Reynolds}, {Stone}, \&
  {Beckwith}}]{2012ApJ...749..189S}
{Sorathia}, K.~A., {Reynolds}, C.~S., {Stone}, J.~M., \& {Beckwith}, K. 2012,
  \apj, 749, 189

\bibitem[{{Tout} \& {Pringle}(1992)}]{1992MNRAS.259..604T}
{Tout}, C.~A., \& {Pringle}, J.~E. 1992, \mnras, 259, 604

\bibitem[{{Velikhov}(1959)}]{Velikhov59}
{Velikhov}, E.~P. 1959, Sov. Phys. JETP, 36

\bibitem[{{Vishniac} \& {Cho}(2001)}]{2001ApJ...550..752V}
{Vishniac}, E.~T., \& {Cho}, J. 2001, \apj, 550, 752

\bibitem[{{Walker} \& {Boldyrev}(2017)}]{2017arXiv170408636W}
{Walker}, J., \& {Boldyrev}, S. 2017, ArXiv e-prints, arXiv:1704.08636

\bibitem[{{Zhou} {et~al.}(2010){Zhou}, {Zhang}, {Wang}, \&
  {Zhu}}]{2010ApJ...710...16Z}
{Zhou}, X.-L., {Zhang}, S.-N., {Wang}, D.-X., \& {Zhu}, L. 2010, \apj, 710, 16

\bibitem[{{Ziegler} \& {R{\"u}diger}(2001)}]{2001A&A...378..668Z}
{Ziegler}, U., \& {R{\"u}diger}, G. 2001, \aap, 378, 668

\end{thebibliography}
\end{document}